\documentclass[11pt]{article}
\usepackage{amsfonts,amsthm,amssymb}
\usepackage[pdftex]{graphicx}
\usepackage{color}
\usepackage[fleqn]{amsmath}
\usepackage[noadjust]{cite}
\usepackage{enumitem}
\usepackage{comment}
\usepackage[leftcaption]{sidecap}
\newcommand{\be}{\begin{eqnarray}}
\newcommand{\ee}{\end{eqnarray}}
\newcommand{\bez}{\begin{eqnarray*}}
\newcommand{\eez}{\end{eqnarray*}}
\newcommand{\pa}{\partial}
\newcommand{\cA}{\mathcal{A}}

\newcommand{\cE}{\mathcal{E}}

\newcommand{\bbC}{\mathbb{C}}

\newcommand{\bbR}{\mathbb{R}}
\newcommand{\bbS}{\mathbb{S}}
\newcommand{\bbZ}{\mathbb{Z}}

\renewcommand{\d}{\mathrm{d}}
\newcommand{\bd}{\bar{\mathrm{d}}}

\newcommand{\imag}{\mathrm{i}}

\newcommand{\bsy}{\boldsymbol}

\theoremstyle{plain}
\newtheorem{theorem}{Theorem}[section]

\theoremstyle{definition}

\newtheorem{remark}[theorem]{Remark}
\newtheorem{example}[theorem]{Example}

\numberwithin{equation}{section}
\numberwithin{theorem}{section}

\renewcommand{\theequation} {\arabic{section}.\arabic{equation}}

\begin{document}

\title{Self-Consistent Sources for Integrable Equations   \\ 
  via Deformations of Binary Darboux Transformations}

\author{
\sc{O. Chvartatskyi}$^{a,b}$, \sc{A. Dimakis}$^c$ and \sc{F. M\"uller-Hoissen}$^b$ \\
 \small
 $^a$ Mathematisches Institut,
 Georg-August Universit\"at G\"ottingen, Bunsenstr. 3-5, 37073 G\"ottingen, Germany \\
  \small
 $^b$ Max-Planck-Institute for Dynamics and Self-Organization,
         37077 G\"ottingen, Germany \\
  \small
 $^c$ Dept. of Financial and Management Engineering,
 University of the Aegean, 82100 Chios, Greece
}

\date{}

\maketitle


\begin{abstract}
We reveal the origin and structure of self-consistent source extensions of integrable equations from 
the perspective of binary Darboux transformations. They arise via a deformation of the potential 
that is central in this method. As examples, we obtain in particular matrix versions of self-consistent 
source extensions of the KdV, Boussinesq, sine-Gordon, nonlinear Schr\"odinger, KP, Davey-Stewartson, two-dimensional 
Toda lattice and discrete KP equation. We also recover a (2+1)-dimensional version of the Yajima-Oikawa system from 
a deformation of the pKP hierarchy. By construction, these systems are accompanied by a hetero binary Darboux transformation, 
which generates solutions of such a system from a solution of the source-free system and additionally solutions 
of an associated linear system and its adjoint. 
The essence of all this is encoded in universal equations in the framework of bidifferential calculus. 
\end{abstract}

\section{Introduction}
The study of soliton equations with self-consistent sources has been pursued in particular in the work of 
Mel'nikov \cite{Mel'83,Mel'84,Mel'88,Mel'89CMPa,Mel'89CMPb,Mel'92}. Mathematically, such systems of equations arise via a multiscaling 
limit of familiar integrable systems (see, e.g., \cite{Zakh+Kuzn86,Mel'89CMPb}), or via a symmetry constraint 
imposed on a higher than two-dimensional integrable system (see \cite{KSS91,Oeve93}, and also, e.g., 
\cite{Kric95,ANP96,SSS99,Helm+Leur07} for related work). Several of these systems appeared, 
independently from those more mathematical explorations, in various physical contexts. For example, 
the nonlinear Schr\"odinger (NLS) equation with 
self-consistent sources describes the nonlinear interaction of an electrostatic high-frequency wave with 
the ion acoustic wave in a plasma (cold ions, warm electrons) in a small amplitude limit 
\cite{Lati+Leon91,CLL91}. In nonlinear optics it describes the interaction of self-induced transparency 
and NLS solitons \cite{Dokt+Vlas83,NYK91}. 
By now quite a number of publications have been devoted to such equations.  

In this work, we show that self-consistent source extensions arise via a simple deformation of the ``potential'' 
that appears in the binary Darboux transformation method (see, e.g., \cite{Matv+Sall91}).
Moreover, the present work provides a more universal approach to such extensions, generalizations 
to matrix versions of such equations, and a corresponding solution generating method. This is achieved in 
the framework of bidifferential calculus \cite{DMH00a,DMH08bidiff,DMH13SIGMA}. 

In Section~\ref{sec:intro_KP_scs} we consider the example of the potential KP (pKP) equation 
with self-consistent sources. The underlying structure is then abstracted to 
bidifferential calculus in Section~\ref{sec:scs_via_bidiff}. The resulting system allows to generate 
self-consistent source extensions of other integrable equations and supplies them with a solution-generating 
method. Examples are treated in the further sections. 
In Appendix~\ref{app:pKPhier}, we apply the aforementioned deformation to the matrix pKP hierarchy. The first 
non-trivial member then turns out to be a (2+1)-dimensional version \cite{Mel'83} 
of the Yajima-Oikawa system \cite{Yaji+Oika76}. 
Finally, Section~\ref{sec:conclusions} contains some concluding remarks.

\section{Via binary Darboux transformation to self-consistent source extensions of the pKP equation}
\label{sec:intro_KP_scs}
\setcounter{equation}{0}
Let $\phi_0$ solve the matrix potential KP (pKP) equation, i.e.,
\be
 \left( 4 \, \phi_{0,t} - \phi_{0,xxx} - 6 \, (\phi_{0,x})^2 \right)_x
       - 3 \, \phi_{0,yy} + 6 \, [\phi_{0,x},\phi_{0,y}] = 0  \, ,   \label{pKP}
\ee
where a subscript $x,y$ or $t$ indicates a partial derivative with respect to this variable. A comma is used to separate 
them from subscripts of a different kind. 
Let $\theta, \eta$ solve the associated linear system and its adjoint, i.e., 
\be
   \theta_y = \theta_{xx} + 2 \, \phi_{0,x} \, \theta \, , \qquad
   \theta_t = \theta_{xxx} + 3 \, \phi_{0,x} \, \theta_x + \frac{3}{2} \left( \phi_{0,y} + \phi_{0,xx} \right) \, \theta 
               \label{pKP_linsys}
\ee   
and             
\be
   \eta_y = - \eta_{xx} - 2 \, \eta \, \phi_{0,x} \, , \qquad
   \eta_t = \eta_{xxx} + 3 \, \eta_x \, \phi_{0,x} - \frac{3}{2} \, \eta \, \left( \phi_{0,y} - \phi_{0,xx} \right) \, .
            \label{pKP_adjlinsys}
\ee
Both pairs of equations have the above pKP equation as consistency condition. As a consequence of these equations, 
the compatibility conditions of the system 
\bez
   \Omega_x = - \eta \, \theta \, , \qquad
   \Omega_y = - \eta \, \theta_x + \eta_x \, \theta  \, , \qquad
   \Omega_t = - \eta \, \theta_{xx} + \eta_x \, \theta_x 
         - \eta_{xx} \, \theta - 3 \, \eta \, \phi_{0,x} \, \theta \, ,
\eez
are satisfied, which guarantees the existence of a ``potential'' $\Omega$. Then it follows that
\be
     \phi = \phi_0 - \theta \, \Omega^{-1} \, \eta    \label{pKP_phi}
\ee
is a new solution of (\ref{pKP}), and 
\be
     q = \theta \, \Omega^{-1} \, , \qquad
     r = \Omega^{-1} \, \eta \, ,     \label{pKP_q,r}
\ee
solve (\ref{pKP_linsys}) and (\ref{pKP_adjlinsys}), respectively, with $\phi$ instead of $\phi_0$. This 
is an essential step of the binary Darboux transformation method for the pKP equation.

\begin{remark} 
An attempt to iterate 
this procedure by using $\phi$, $q$ and $r$ instead of $\phi_0$, $\theta$ and $\eta$, leads back to the $\phi_0$ we 
started with to generate $\phi$. Iteration of the binary Darboux transformation method involves the construction 
of still other solutions of the linear systems. But we will not need these further steps here, since 
we consider a \emph{vectorial} binary Darboux transformation, so that there is no need for iteration. 
\end{remark}

Let us now deform the potential $\Omega$ as follows, 
\be
   \Omega_x &=& - \eta \, \theta + c_1 \, , \nonumber \\
   \Omega_y &=& - \eta \, \theta_x + \eta_x \, \theta + c_2  \, , \nonumber \\
   \Omega_t &=& - \eta \, \theta_{xx} + \eta_x \, \theta_x 
         - \eta_{xx} \, \theta - 3 \, \eta \, \phi_{0,x} \, \theta + c_3  \, ,   \label{pKP_Omega_eqs}
\ee
with (matrix) functions $c_i$, $i=1,2,3$. Consistency requires that
\bez
     c_1 = \omega_x \, , \qquad
     c_2 = \omega_y \, , \qquad
     c_3 = \omega_t \, ,
\eez
with a potential $\omega$. Hence the above deformation actually amounts to the substitution 
$\Omega \mapsto \Omega - \omega$ in the previous equations. Then $\phi,q,r$ no longer satisfy (\ref{pKP}), 
(\ref{pKP_linsys}) and (\ref{pKP_adjlinsys}). Instead we find
\be
   q_y - q_{xx} &=& 2 \, \phi_x \, q 
       + \left( - q \, \omega_y + q \, \omega_{xx} + 2 \, q_x \, \omega_x \right) \, \Omega^{-1} \, , 
             \nonumber \\
   q_t - q_{xxx} &=& 3 \, \phi_x \, q_x 
       + \frac{3}{2} \left( \phi_y + \phi_{xx} \right) \, q 
       + \frac{3}{2} \, \left( - q \, \omega_y + q \, \omega_{xx} + 2 \, q_x \, \omega_x \right) \, r \, q \nonumber \\
       && + \left( - q \, \omega_t + q \, \omega_{xxx} + 3 \, (q_x \, \omega_x)_x 
                  + 3 \, \phi_x \, q \, \omega_x \right) \, \Omega^{-1} \, , \label{pKP_q-sys_scs} \\
   r_y + r_{xx} &=& - 2 \, r \, \phi_x 
           - \Omega^{-1} \left( \omega_y \, r + \omega_{xx} \, r + 2 \, \omega_x \, r_x \right) \, , \nonumber \\
   r_t - r_{xxx} &=& 3 \, r_x \, \phi_x - \frac{3}{2} \, r \, \left( \phi_y - \phi_{xx} \right)
        + \frac{3}{2} \, r \, q \, \left( \omega_y \, r + \omega_{xx} \, r + 2 \, \omega_x \, r_x \right) \nonumber \\
       && + \Omega^{-1} \, \left( - \omega_t \, r + \omega_{xxx} \, r + 3 \, (\omega_x \, r_x)_x 
          + 3 \, \omega_x \, r \, \phi_x \right) \, ,   \label{pKP_r-sys_scs}    
\ee   
and the extended pKP equation     
\be
  \lefteqn{ \left( 4 \, \phi_t - \phi_{xxx} - 6 \, (\phi_x)^2 \right)_x
       - 3 \, \phi_{yy} + 6 \, [\phi_x,\phi_y] } \hspace{1cm}
       \nonumber \\
  &=& \left( 4 \, q \, \omega_t \, r - 6 \, q \, \omega_x \, r_y - 6 \, q \, \omega_x \, r_{xx} 
      + 6 \, q \, \omega_y \, r_x - 18 \, q \, \omega_{xx} \, r_x - 6 \, q \, \omega_{xy} \, r 
      - 10 \, q \, \omega_{xxx} \, r \right)_x \nonumber \\
  && + \left( 6 \, q \, \omega_x \, r_x - 3 \, q \, \omega_y \, r + 9 \, q \, \omega_{xx} \, r \right)_y 
     + \left( 6 \, q \, \omega_x \, r_x - 3 \, q \, \omega_y \, r + 9 \, q \, \omega_{xx} \, r \right)_{xx}
       \, .   \label{pKP_scs}
\ee
If $\phi$ is an $m \times m$ matrix, then $\theta,q$ have matrix size $m \times n$, and 
$\eta,r$ have size $n \times m$. $\Omega$ and $\omega$ are then $n \times n$ matrices.

\begin{remark}
\label{rem:pKP_inverse_bDT}
Using (\ref{pKP_q,r}), and introducing $\hat{\Omega} := - \Omega^{-1}$, we turn (\ref{pKP_Omega_eqs}) into the Riccati system
\bez
  &&  \hat{\Omega}_x = - r \, q + \hat{\Omega} \, \omega_x \, \hat{\Omega}\, , \qquad
      \hat{\Omega}_y = - r \, q_x + r_x \, q + \hat{\Omega} \, \omega_y \, \hat{\Omega} \, , \\
  &&  \hat{\Omega}_t = - r \, q_{xx} + r_x \, q_x - r_{xx} \, q - 3 \, r \, \phi_x \, q + \hat{\Omega} \, \omega_t \, \hat{\Omega} \, .
\eez
This completes the system consisting of (\ref{pKP_q-sys_scs}), (\ref{pKP_r-sys_scs}) (where $\Omega^{-1}$ should 
be replaced by $-\hat{\Omega}$) and (\ref{pKP_scs}). 
The above hetero binary Darboux transformation can now be inverted to map solutions of (\ref{pKP_q-sys_scs}), (\ref{pKP_r-sys_scs}) 
and (\ref{pKP_scs}) to solutions of the source-free pKP equation and solutions of its linear system:
\bez
     \phi_0 = \phi - q \, \hat{\Omega}^{-1} r \, , \quad
     \theta = - q \, \hat{\Omega}^{-1} \, , \quad
     \eta = - \hat{\Omega}^{-1} \, r  \, .
\eez
The equations for $q,r$ and $\hat{\Omega}$ are \emph{non}linear, however, and thus more difficult to solve. 
\end{remark}

Next we look for choices of $\omega$, such that $\Omega$ can be eliminated in half of 
the equations (\ref{pKP_q-sys_scs}) and (\ref{pKP_r-sys_scs}). This requires $\omega_x=0$ 
and leaves us with the choices $\omega_y =0$ or $\omega_t=0$. In the first case, we only keep the 
equations for $q_y$ and $r_y$. In the second case, we only keep those for $q_t$ and $r_t$. 
This results in the two versions of pKP with self-consistent sources that appeared in the 
literature. 
\begin{enumerate}
\item $\omega = \omega(t)$.
\be
 && \left( 4 \, \phi_t - \phi_{xxx} - 6 \, (\phi_x)^2 \right)_x
       - 3 \, \phi_{yy} + 6 \, [\phi_x,\phi_y] 
  = 4 \, \left( q \, \omega_t \, r \right)_x \, ,   \nonumber     \\
 && q_y - q_{xx} = 2 \, \phi_x \, q  \, , \qquad
    r_y + r_{xx} = - 2 \, r \, \phi_x \, .   \label{pKP_scs_1_mod}  
\ee
In terms of 
\be
     \tilde{q} = q \, \mathcal{Q} \, , \qquad
     \tilde{r} = \mathcal{R} \, r \, ,    \label{q,r_transformation}
\ee        
with a suitable choice of $\mathcal{Q}(t)$ and $\mathcal{R}(t)$, this system can be written as  
\be
 && \left( 4 \, \phi_t - \phi_{xxx} - 6 \, (\phi_x)^2 \right)_x
       - 3 \, \phi_{yy} + 6 \, [\phi_x,\phi_y] 
  = (\tilde{q} \, \tilde{r})_x \, ,   \nonumber     \\
 && \tilde{q}_y - \tilde{q}_{xx} = 2 \, \phi_x \, \tilde{q}  \, , \qquad
    \tilde{r}_y + \tilde{r}_{xx} = - 2 \, \tilde{r} \, \phi_x \, ,    \label{pKP_scs_1}  
\ee
where $\omega_t$ has been absorbed. 
The latter is rather what we should call a system with self-consistent sources. 
Its scalar version has been studied in 
\cite{Mel'83,Mel'87,Mel'89CMPa,Zakh+Kuzn86,DCZ03,Xiao+Zeng04,LZL08KP,LLZ14}. The ``noncommutative'' generalization 
(\ref{pKP_scs_1}) already appeared in \cite{Chva+Sydo15}. 
In our framework, the modification (\ref{pKP_scs_1_mod}) is important. By actually solving (\ref{pKP_scs_1_mod}), 
we obtain solutions of the self-consistent source system (\ref{pKP_scs_1}), which depend on arbitrary functions of $t$. 
The appearance of arbitrary functions of a single variable in solutions is a generic feature of systems with 
self-consistent sources. 
\item $\omega=\omega(y)$.
\be
  && \left( 4 \, \phi_t - \phi_{xxx} - 6 \, (\phi_x)^2 \right)_x
       - 3 \, \phi_{yy} + 6 \, [\phi_x,\phi_y] 
  = 3 \, \left( q \, \omega_y \, r_x - q_x \, \omega_y \, r \right)_x 
     - 3 \,  \left( q \, \omega_y \, r \right)_y   \, ,  \qquad \quad  \nonumber   \\
  && q_t - q_{xxx} = 3 \, \phi_x \, q_x 
       + \frac{3}{2} \left( \phi_y + \phi_{xx} \right) \, q 
       - \frac{3}{2} \, q \, \omega_y \, r \, q  \, , \nonumber \\
  && r_t - r_{xxx} = 3 \, r_x \, \phi_x - \frac{3}{2} \, r \, \left( \phi_y - \phi_{xx} \right)
        + \frac{3}{2} \, r \, q \, \omega_y \, r \, .   \label{pKP_scs_2_mod} 
\ee
Again, introducing new variables as in (\ref{q,r_transformation}), now with a suitable $\mathcal{Q}(y)$ and $\mathcal{R}(y)$, 
$\omega_y$ can be absorbed. The scalar version of this system has been studied in \cite{Mel'83,LZL08KP}.
\end{enumerate}

\begin{remark}
In the scalar case ($m=1$), if $\omega$ does not depend on $x$, the expression (\ref{pKP_phi}) can be 
rewritten as $\phi = \phi_0 + \mathrm{tr}( - \theta \, \Omega^{-1} \, \eta) = \phi_0 + \mathrm{tr}( \Omega_x \, \Omega^{-1}) 
 = \phi_0 + (\ln(\det \Omega))_x$.
Together with (\ref{pKP_q,r}), this provides us with solutions of the scalar versions of 
(\ref{pKP_q-sys_scs}) and (\ref{pKP_r-sys_scs}). If $\omega$ depends only on $t$ or $y$, 
we obtain solutions of (\ref{pKP_scs_1_mod}) and (\ref{pKP_scs_2_mod}), respectively.
\end{remark}

\begin{remark}
As formulated above, the number of sources appears to be $n$. However, this is only so if $\omega_t$, respectively 
$\omega_y$, has maximal rank. If the rank is $N < n$, then only $N$ sources appear on the right hand side 
of (\ref{pKP_scs}). 
\end{remark}

The above procedure provides us with a hetero binary Darboux transformation from the pKP equation and its 
associated linear system to any of the pKP systems with self-consistent sources (modified by $\omega$).\footnote{In 
the scalar case, hetero binary Darboux transformations from a KP equation with self-consistent sources to the 
KP equation with additional sources have been elaborated in \cite{Xiao+Zeng04}. }
In the framework of bidifferential calculus, we can abstract the underlying structure from the specific example 
(here pKP) and then obtain corresponding self-consistent source extensions of quite a number of other 
integrable equations.  

Remark~\ref{rem:pKP_inverse_bDT} shows that there is also a transformation that maps a class of solutions of 
any of the self-consistent source extensions of the pKP equation to solutions of the source-free pKP equation 
and its linear system. We expect that this is a general feature of integrable systems with self-consistent sources.

\paragraph{Exact solutions in case of constant seed.}
Let $\phi_0$ be constant. Special solutions of (\ref{pKP_linsys}) and (\ref{pKP_adjlinsys}) 
are given by 
\bez
   \theta = a \, e^{\vartheta(P)} \, A \, , \qquad 
   \eta = B \, e^{-\vartheta(Q)} \, b \, , \qquad
   \vartheta(P) := P \, x + P^2 \, y + P^3 \, t  \, , 
\eez
with constant matrices $a$, $b$, $A$, $B$, $P$ and $Q$ of appropriate size. Then (\ref{pKP_Omega_eqs}) is solved by
\bez
    \Omega = \omega - B \, e^{-\vartheta(Q)} \, X \, e^{\vartheta(P)} \, A  \, ,
\eez
with a constant matrix $X$ that satisfies the Sylvester equation
\bez
    X \, P - Q \, X  = b \, a  \, .
\eez
(\ref{pKP_phi}) and (\ref{pKP_q,r}) now provide us with explicit solutions of the above matrix 
pKP equations with self-consistent sources, (\ref{pKP_scs_1_mod}) and (\ref{pKP_scs_2_mod}). The basic soliton solutions 
are obtained if $P$ and $Q$ are diagonal with distinct eigenvalues, and if they have disjoint spectra, in which 
case the Sylvester equation has a unique solution. Also see \cite{Sakh03} for the source-free case. 

\begin{example}
In the scalar case ($m=1$), choosing $n=1$, $a=A=B=b=1$ and 
$\omega = e^{\alpha}/(Q-P)$, where $\alpha$ is either a function only of $y$ or of $t$, 
we obtain 
\bez
 && \phi = \phi_0 + \frac{(P-Q) \, e^{{\vartheta(P)}-\vartheta(Q)}}{e^{\alpha}+e^{{\vartheta(P)}-\vartheta(Q)}} 
       = \phi_0 + \frac{1}{2} (P-Q) + \left( \ln (\cosh \Theta) \right)_x \, ,  \\
 && q = \frac{(Q-P) \, e^{{\vartheta(P)}}}{e^{\alpha}+e^{{\vartheta(P)}-\vartheta(Q)}} 
      = \frac{1}{2} (Q-P) \, e^{\tilde{\Theta}} \, \mathrm{sech} (\Theta) \, , \;
     r = \frac{(Q-P) \, e^{-{\vartheta(Q)}}}{e^{\alpha}+e^{{\vartheta(P)}-\vartheta(Q)}} 
       = \frac{1}{2} (Q-P) \, e^{-\tilde{\Theta}-\alpha} \, \mathrm{sech}(\Theta) ,
\eez 
where
$\Theta = \frac{1}{2} \left(\vartheta(P)-\vartheta(Q)-\alpha\right)$ and
$\tilde{\Theta} = \frac{1}{2} \left(\vartheta(P) + \vartheta(Q) - \alpha \right)$. 
Then $u = \phi_x$, together with $q$ and $r$, or rather the transformed $\tilde{q}$ and $\tilde{r}$, is a soliton solution
of the scalar KP equation with self-consistent sources. Also see \cite{HHOS89,DCZ03,Xiao+Zeng04,KRL04}. 
\end{example}

\section{A framework for generating self-consistent source extensions of integrable equations}
\label{sec:scs_via_bidiff}
\setcounter{equation}{0}
Let us recall some basics of bidifferential calculus. 
A \emph{graded associative algebra} is an associative algebra $\boldsymbol{\Omega} = \bigoplus_{r \geq 0} \boldsymbol{\Omega}^r$
over $\bbC$, where $\cA := \boldsymbol{\Omega}^0$ is an associative algebra over $\bbC$ 
and $\boldsymbol{\Omega}^r$, $r \geq 1$,
are $\cA$-bimodules such that $\boldsymbol{\Omega}^r \, \boldsymbol{\Omega}^s \subseteq \boldsymbol{\Omega}^{r+s}$.
A \emph{bidifferential calculus} is a unital graded associative algebra $\boldsymbol{\Omega}$, supplied
with two ($\bbC$-linear) graded derivations $\d, \bar{\d} : \boldsymbol{\Omega} \rightarrow \boldsymbol{\Omega}$
of degree one (hence $\d \boldsymbol{\Omega}^r \subseteq \boldsymbol{\Omega}^{r+1}$,
$\bar{\d} \boldsymbol{\Omega}^r \subseteq \boldsymbol{\Omega}^{r+1}$), and such that
\be
    \d^2 = \bd^2 = \d \bd + \bd \d = 0  \, .   \label{bidiff_conds}
\ee
In this framework, many integrable equations can be expressed either as
\be
     \d \, \bd \, \phi_0 + \d\phi_0 \; \d\phi_0 = 0 \, ,    \label{phi_eq}
\ee
with $\phi_0 \in \mathrm{Mat}(m,m,\cA)$ (the algebra of $m \times m$ matrices over $\cA$), 
and possibly with some reduction condition, or as
\be
        \d \, [ (\bd g_0) \, g_0^{-1} ] = 0  \, ,   \label{g_eq}
\ee
with an invertible $g_0 \in \mathrm{Mat}(m,m,\cA)$. The two equations are related by the \emph{Miura equation}
\be
     \bd g_0  = (\d \phi_0) \, g_0 ,  \label{Miura}
\ee 
which has both, (\ref{phi_eq}) and (\ref{g_eq}), as integrability conditions. (\ref{phi_eq}) and (\ref{g_eq}) 
are generalizations or analogs of well-known potential forms of the self-dual Yang-Mills equation 
(cf. \cite{DMH08bidiff}). 
\vspace{.3cm}

A linear system and an adjoint linear system for (\ref{phi_eq}) is given by 
\be
    \bd \theta = (\d \phi_0) \, \theta + (\d \theta) \, \Delta + \theta \, \lambda \, ,  \label{theta_eq}         
\ee
respectively
\be
    \bd \eta = - \eta \, \d \phi_0 + \Gamma \, \d \eta + \kappa \, \eta \, ,   \label{eta_eq}
\ee
where $\theta \in \mathrm{Mat}(m,n,\cA)$, $\eta \in \mathrm{Mat}(n,m,\cA)$,  
$\Delta,\Gamma \in \mathrm{Mat}(n,n,\cA)$, and $\lambda,\kappa$ are $n \times n$ matrices of
elements of $\Omega^1$. They have to satisfy\footnote{For $\kappa=\lambda=0$, these are ``Riemann 
equations in bidifferential calculus'' \cite{CMHS15}.}
\be
  &&  \bd \Delta + [\lambda , \Delta] = (\d \Delta) \, \Delta \, , \qquad  
      \bd \lambda + \lambda^2 = (\d \lambda) \, \Delta \, , \nonumber \\ 
  &&  \bd \Gamma - [\kappa , \Gamma] = \Gamma \, \d \Gamma  \, , \qquad  
      \bd \kappa - \kappa^2 = \Gamma \, \d \kappa \, ,   \label{Delta,lambda,Gamma,kappa_eqs}
\ee
as a consequence of (\ref{bidiff_conds}) and the graded derivation property of $\d$ and $\bd$.

\paragraph{Binary Darboux transformation.}
Let $\Omega$ be a solution of
\be
  &&  \Gamma \, \Omega - \Omega \, \Delta = \eta \, \theta \, ,   \label{Omega_Sylvester} \\
  &&  \bd \Omega = (\d \Omega) \, \Delta - (\d \Gamma) \, \Omega 
     + \kappa \, \Omega + \Omega \, \lambda + (\d \eta) \, \theta \, .  \label{bd_Omega} 
\ee
The equation obtained by acting with $\bd$ on (\ref{Omega_Sylvester}) 
is identically satisfied as a consequence of (\ref{theta_eq}), (\ref{eta_eq}), (\ref{Omega_Sylvester}), 
and the equation that results from (\ref{Omega_Sylvester}) by acting with $\d$ on it.   
Correspondingly, also the equation that results from acting with $\bd$ on (\ref{bd_Omega}) is identically satisfied 
as a consequence of the preceding equations. It follows \cite{DMH08bidiff} that
\bez
   \phi = \phi_0 - \theta \, \Omega^{-1} \, \eta 
\eez
is a new solution of (\ref{phi_eq}), and 
\bez  
   q = \theta \, \Omega^{-1} \, , \qquad 
   r = \Omega^{-1} \, \eta \, ,
\eez
satisfy 
\bez
    \bd q = (\d \phi) \, q + \d( q \, \Gamma) - q \, \kappa  \, ,  \qquad
    \bd r = - r \, (\d \phi) + \d( \Delta \, r) - \lambda \, r  \, .       
\eez

\paragraph{Deformation of the potential.}
Guided by the pKP example in the preceding section, we replace $\Omega$ by $\Omega - \omega$ in 
the above equations, i.e., 
\bez
  &&  \Gamma \, (\Omega - \omega) - (\Omega - \omega) \, \Delta = \eta \, \theta \, , \\
  &&  \bd (\Omega - \omega) = \d (\Omega - \omega) \, \Delta - (\d \Gamma) \, (\Omega - \omega) 
     + \kappa \, (\Omega - \omega) + (\Omega - \omega) \, \lambda + (\d \eta) \, \theta  \, .
\eez
Hence
\be
  &&  \Gamma \, \Omega - \Omega \, \Delta = \eta \, \theta + c \, , \nonumber  \\
  &&  \bd \Omega = (\d \Omega) \, \Delta - (\d \Gamma) \, \Omega 
     + \kappa \, \Omega + \Omega \, \lambda + (\d \eta) \, \theta + \gamma \, ,   \label{scs_Omega_eqs}
\ee
where
\be
     c := \Gamma \, \omega - \omega \, \Delta \, , \qquad
    \gamma := \bd \omega - (\d \omega) \, \Delta + (\d \Gamma) \, \omega - \kappa \, \omega 
              - \omega \, \lambda \, .
                                   \label{c,gamma_in_terms_of_omega}
\ee
We note that they satisfy 
\be
   && \bd \gamma = (\d \gamma) \, \Delta - (\d \Gamma) \, \gamma + \kappa \, \gamma - \gamma \, \lambda 
      - (\d \kappa) \, c \, , \nonumber \\
   && \bd c = (\d c) \, \Delta + \kappa \, c + c \, \lambda + \Gamma \, \gamma - \gamma \, \Delta \, .   
       \label{gamma,c_constraints}
\ee
By straightforward computations, one proves the following. 

\begin{theorem}
Let $\phi_0$ be a solution of (\ref{phi_eq}) and let 
$\theta$, $\eta$, $\Omega$ satisfy the linear equations (\ref{theta_eq}), (\ref{eta_eq}) and (\ref{scs_Omega_eqs}), 
respectively.\footnote{Here we also assume, of course, that $\Delta, \Gamma, \kappa, \lambda$ satisfy 
(\ref{Delta,lambda,Gamma,kappa_eqs}). }
Then
\be
    \phi = \phi_0 - \theta \, \Omega^{-1} \, \eta \, , \qquad
       q = \theta \, \Omega^{-1} \, , \qquad 
       r = \Omega^{-1} \, \eta  \, ,  \label{phi,q,r}
\ee
are solutions of
\be
    \d \, \bd \, \phi + \d \phi \; \d \phi = \d ( q \, \gamma \, r - q \, \d (c \, r) )     
                \label{scs_phi_eq}
\ee
and 
\be
    \bd q &=& (\d \phi) \, q + \d( q \, \Gamma) - q \, \kappa 
        - q \, \gamma \, \Omega^{-1} - (\d q) \, c \, \Omega^{-1} \, ,  \nonumber \\
    \bd r &=& - r \, \d \phi + \d( \Delta \, r) - \lambda \, r  
        - \Omega^{-1} \, \gamma \, r + \Omega^{-1} \, \d(c \, r) \, .       \label{scs_q,r_eqs}
\ee
\end{theorem}

 From (\ref{scs_phi_eq}) and (\ref{scs_q,r_eqs}) we can recover the self-consistent source extensions 
of the pKP equation, revisited in the preceding section, see below. 
But now we can choose different bidifferential calculi and obtain self-consistent source extensions also of 
other integrable equations. A number of examples will be presented in the following sections.
In all these examples, we have $c=0$, i.e., 
\be
        \Gamma \, \omega = \omega \, \Delta \, .   \label{G.o=o.D}
\ee   

\begin{remark}
The above equations form a consistent system in the sense that any equation derived from it by acting with 
$\bd$ on any of its members yields an equation that is satisfied as a consequence of the system. 
\end{remark}

\begin{remark}
\label{rem:phi,q,r,hatOmega_sys}
Using the last two equations in (\ref{phi,q,r}), we eliminate $\theta$ and $\eta$ in (\ref{scs_Omega_eqs}) 
to obtain
\bez
  &&  \Delta \, \hat{\Omega} - \hat{\Omega} \, \Gamma = r \, q + \hat{\Omega} \, c \, \hat{\Omega} \, , \\
  &&  \bd \hat{\Omega} = \d (\hat{\Omega} \Gamma) - \hat{\Omega} \, \kappa - \lambda \, \hat{\Omega} + (\d r) \, q
                        + \left( (\d \hat{\Omega}) \, c + \hat{\Omega} \, \gamma \right) \, \hat{\Omega} \, ,
\eez
where $\hat{\Omega} = - \Omega^{-1}$. (\ref{scs_q,r_eqs}) then reads
\bez
    \bd q &=& (\d \phi) \, q + \d( q \, \Gamma) - q \, \kappa 
        + q \, \gamma \, \hat{\Omega} + (\d q) \, c \, \hat{\Omega} \, ,  \nonumber \\
    \bd r &=& - r \, \d \phi + \d( \Delta \, r) - \lambda \, r  
        + \hat{\Omega} \, \gamma \, r - \hat{\Omega} \, \d(c \, r) \, . 
\eez
Together with (\ref{scs_phi_eq}), this constitutes a system of equations for $\phi$, $q$, $r$ and $\hat{\Omega}$, for 
which we now have a solution-generating technique at hand. Also see Remark~\ref{rem:pKP_inverse_bDT}.
\end{remark}

\paragraph{Generating solutions of the equations for $g$.}
If $\phi_0$ and $g_0$ satisfy the Miura equation (\ref{Miura}), 
then $\phi,q,r$ together with
\bez
     g = (I - \theta \, \Omega^{-1} \, \Gamma^{-1} \eta ) \, g_0 \, ,
\eez
where $I$ is the identity matrix, satisfy 
\bez
   \bd g - (\d \phi) \, g = ( q \, \gamma + (\d q) \, c ) \, \Omega^{-1} \, \Gamma^{-1} \, \Omega \, r \, g_0 \, . 
\eez
If (\ref{G.o=o.D}) holds, i.e., $c=0$, then we have
\bez    
     g^{-1} = g_0^{-1} \, (I + \theta \, \Delta^{-1} \, \Omega^{-1} \, \eta ) \, ,
\eez
and $g$ satisfies the \emph{extended Miura equation}
\be
   \bd g - (\d \phi) \, g = ( q \, \gamma \, \Delta^{-1} \, r ) \, g \, ,  \label{ext_Miura}
\ee
which implies the following extension of (\ref{g_eq}), 
\be
     \d [ (\bd g) \, g^{-1} ] = \d (q \, \gamma \, \Delta^{-1} \, r ) \, .  \label{scs_g_eq}
\ee
Via the extended Miura equation (\ref{ext_Miura}), we can eliminate $\phi$ in (\ref{scs_q,r_eqs}) in favor of $g$. 
The resulting equations, together with (\ref{scs_g_eq}), constitute the (Miura-) dual of (\ref{scs_phi_eq})
and (\ref{scs_q,r_eqs}). It is another generating system for further self-consistent source extensions of 
integrable equations.

\paragraph{Transformations.}
The equations for $\Delta$ and $\lambda$ in (\ref{Delta,lambda,Gamma,kappa_eqs}) and the 
linear equation (\ref{theta_eq}) are invariant under a transformation
\bez
    \theta \mapsto \theta \, L \, , \quad
    \Delta \mapsto L^{-1} \, \Delta \, L \, , \quad
    \lambda \mapsto L^{-1} \, \lambda \, L + L^{-1} \, \bd L + (\d L^{-1}) \, \Delta \, L \, , 
\eez
with an invertible $L$. Correspondingly, the equations for $\Gamma$ and $\kappa$ 
in (\ref{Delta,lambda,Gamma,kappa_eqs}) and the linear equation (\ref{eta_eq}) are invariant under
\bez
     \eta \mapsto M \, \eta \, , \quad
     \Gamma \mapsto M \, \Gamma \, M^{-1} \, , \quad
     \kappa \mapsto M \, \kappa \, M^{-1} + (\bd M) \, M^{-1} + M \, \Gamma \, \d M^{-1}  \, ,
\eez
with an invertible $M$. 
(\ref{scs_Omega_eqs}) is invariant if we supplement these transformations by
$\Omega \mapsto M \, \Omega \, L$ and $\omega \mapsto M \, \omega \, L$.
Now (\ref{gamma,c_constraints}) implies $c \mapsto M \, c \, L$ and $\gamma \mapsto M \, \gamma \, L$.  
The formula for $\phi$ in (\ref{phi,q,r}) is invariant, and those for $q$ and $r$ imply
$q \mapsto q \, M^{-1}$ and $r \mapsto L^{-1} \, r$.
It follows that (\ref{scs_phi_eq}), (\ref{scs_q,r_eqs}), (\ref{G.o=o.D}), (\ref{ext_Miura}) and (\ref{scs_g_eq}) are invariant. 
This suggests that, via such a transformation, $\kappa$ and $\lambda$ can typically be eliminated. 
However, Sections~\ref{sec:KdV}-\ref{sec:NLS} show that this may not necessarily be so and 
that special choices of $\kappa$ and $\lambda$ can achieve a drastic simplification of the equations.

\paragraph{Choice of the graded algebra.}
In this work, we specify the graded algebra $\boldsymbol{\Omega}$ to be of the form
\be
    \boldsymbol{\Omega} = \cA \otimes \bsy{\Lambda} \, ,  \qquad
    \bsy{\Lambda} = \bigoplus_{i=0}^2 \bsy{\Lambda}^i    \label{Omega_wedge}
\ee
where $\bsy{\Lambda}$ is the exterior (Grassmann) algebra of the vector space $\bbC^2$. 
It is then sufficient to define $\d$ and $\bd$ on $\cA$, since they extend to $\boldsymbol{\Omega}$ 
in a straightforward way, treating the elements of $\bsy{\Lambda}$ as constants. Moreover, 
$\d$ and $\bd$ extend to matrices over $\boldsymbol{\Omega}$. 
We choose a basis $\xi_1,\xi_2$ of $\bsy{\Lambda}^1$.

\paragraph{Recovering the pKP case.} Let $\cA_0$ be the space of smooth complex functions on $\mathbb{R}^3$. 
We extend it to $\cA = \cA_0[\pa]$, where $\pa:=\pa_x$ is the partial differentiation operator 
with respect to $x$. On $\cA$ we define (cf. \cite{DMH08bidiff}) 
\bez
    \d f = [\pa, f] \, \xi_1 + \frac{1}{2} \, [ \pa_y+\pa^2,f] \, \xi_2 \, , \qquad
    \bd f = \frac{1}{2} \, [\pa_y-\pa^2,f] \, \xi_1
            + \frac{1}{3} \, [\pa_t - \pa^3,f] \, \xi_2 \, .   
\eez
The maps $\d$ and $\bd$ extend to linear maps on $\boldsymbol{\Omega}=\cA \otimes \bsy{\Lambda}$, 
and moreover to matrices over $\boldsymbol{\Omega}$. Choosing 
\bez
      \Delta = \Gamma = -I_n \pa \, ,   \qquad     \kappa = \lambda = 0 \, , 
\eez 
where $I_n$ is the $n \times n$ identity matrix, 
(\ref{Delta,lambda,Gamma,kappa_eqs}) is satisfied, and $c=0$, i.e., (\ref{G.o=o.D}), becomes $\omega_x =0$. 
The second equation in (\ref{c,gamma_in_terms_of_omega}) leads to
\bez
      \gamma = \frac{1}{2} \omega_y \, \xi_1 
               + \left( \frac{1}{3} \omega_t + \frac{1}{2} \omega_y \, \pa \right) \, \xi_2 \, .
\eez
(\ref{scs_Omega_eqs}) becomes (\ref{pKP_Omega_eqs}) with $c_1=0$, $c_2=\omega_y$, $c_3=\omega_t$. 
Taking $\omega_x =0$ into account, we recover all equations in Section~\ref{sec:intro_KP_scs}. 
For example, (\ref {scs_q,r_eqs}) yields (\ref{pKP_q-sys_scs}) and (\ref{pKP_r-sys_scs}),   
and (\ref{scs_phi_eq}) becomes (\ref{pKP_scs}). 
A corresponding extension of the whole pKP hierarchy is presented in Appendix~\ref{app:pKPhier}.

\section{Matrix KdV equation with self-consistent sources}
\label{sec:KdV}
\setcounter{equation}{0}
Let $\cA_0$ be the space of smooth complex functions on $\mathbb{R}^2$. 
We extend it to $\cA = \cA_0[\pa]$, where $\pa:=\pa_x$ is the partial differentiation operator 
with respect to the coordinate $x$. On $\cA$ we define (cf. \cite{DMH08bidiff}) 
\bez
    \d f = [\pa, f] \, \xi_1 + \frac{1}{2} \, [\pa^2,f] \, \xi_2 \, , \qquad
    \bd f = -\frac{1}{2} \, [\pa^2,f] \, \xi_1
            + \frac{1}{3} \, [\pa_t - \pa^3,f] \, \xi_2 \, .   
\eez
The maps $\d$ and $\bd$ extend to linear maps on $\boldsymbol{\Omega}=\cA \otimes \bsy{\Lambda}$, 
and moreover to matrices over $\boldsymbol{\Omega}$. Choosing 
\bez
      \Delta = \Gamma = -I_n \pa \, ,   \quad 
      \kappa = \frac{1}{2} \, Q^2 \, (\xi_1 + \pa \, \xi_2) \, ,   \quad 
      \lambda = -\frac{1}{2} \, P^2 \, (\xi_1 + \pa \, \xi_2) \, ,
\eez 
with constant matrices $P,Q$, 
(\ref{Delta,lambda,Gamma,kappa_eqs}) is satisfied, and (\ref{G.o=o.D}) becomes $\omega_x =0$. 
The choices for $\kappa$ and $\lambda$ considerably simplify the subsequent equations.\footnote{$\kappa = \lambda =0$ 
is not the best choice here.}
The second equation in (\ref{c,gamma_in_terms_of_omega}) yields
\bez
      \gamma = \gamma_1 \, \xi_1 + (\gamma_2 + \gamma_1 \, \pa) \, \xi_2 \, , \qquad
      \gamma_1 = - \frac{1}{2} (Q^2 \omega - \omega \, P^2) \, , \quad
      \gamma_2 = \frac{1}{3} \, \omega_t \, .
\eez
The linear equations (\ref{theta_eq}) and (\ref{eta_eq}) read
\bez
  &&  \theta_{xx} = \theta \, P^2- 2 \, \phi_{0,x} \, \theta   \, , \quad
    \theta_t = \theta_x \, P^2 + \phi_{0,x} \, \theta_x - \frac{1}{2} \phi_{0,xx} \, \theta \, , \\
  && \eta_{xx} = Q^2 \eta - 2 \eta \, \phi_{0,x}   \, , \quad
     \eta_t = Q^2 \eta_x + \eta_x \, \phi_{0,x} - \frac{1}{2} \eta \, \phi_{0,xx} \, ,
\eez
and (\ref{scs_Omega_eqs}) takes the form
\be
  && Q^2 \, \Omega - \Omega \, P^2 = \eta \, \theta_x - \eta_x \, \theta + Q^2 \, \omega - \omega \, P^2 \, , \nonumber \\
  && \Omega_x = - \eta \, \theta \, , \quad
     \Omega_t = - Q^2 \, \eta \, \theta - \eta \, \theta \, P^2 + \eta_x \, \theta_x 
              + \eta \, \phi_{0,x} \, \theta + \omega_t \, .   \label{KdV_Omega_eqs}
\ee
According to Section~\ref{sec:scs_via_bidiff}, it follows that $\phi$, $q$ and $r$, 
given by (\ref{phi,q,r}), satisfy 
\be
  &&  4 \, u_t - u_{xxx} - 3 \, (u^2)_x  
     = 8 \, (q \, \omega_t \, r)_x + 6 \, \left( q_x \, (Q^2 \omega - \omega P^2) \, r  
       - q \, (Q^2 \omega - \omega P^2) \, r_x  \right)_x \, , \nonumber  \\
  && q_{xx} = q \, Q^2  - u \, q - q \, (Q^2 \, \omega - \omega P^2) \, \Omega^{-1}  \, , \nonumber\\
  && q_t = q_x \, Q^2 + \frac{1}{2} q \, (Q^2 \omega - \omega \, P^2) \, r \, q
            + \frac{1}{2} u \, q_x - \frac{1}{4} u_x \, q  
            - q_x \, (Q^2 \omega - \omega \, P^2) \, \Omega^{-1} - q \, \omega_t \, \Omega^{-1}   \nonumber\\ 
  && \quad  = q_{xxx} + \frac{3}{2} u \, q_x  + \frac{3}{4} u_{x} \, q 
             + \frac{3}{2} q \, (Q^2 \omega - \omega \, P^2) \, r \, q 
             - q \, \omega_t \, \Omega^{-1} \, , \nonumber\\ 
  && r_{xx} = P^2 \, r - r \, u + \Omega^{-1} \, (Q^2 \omega - \omega \, P^2)  \, ,   \nonumber \\
  && r_t = P^2 \, r_x - \frac{1}{2} r \, q \, (Q^2 \omega - \omega \, P^2) \, r
           + \frac{1}{2} r_x \, u - \frac{1}{4} r \, u_x
           + \Omega^{-1} \, (Q^2 \omega - \omega \, P^2) \, r_x - \Omega^{-1} \, \omega_t \, r  \nonumber \\
  && \quad = r_{xxx} + \frac{3}{2} r_x \, u + \frac{3}{4} r \, u_x - \frac{3}{2} r \, q \, (Q^2 \omega - \omega \, P^2) \, r 
             - \Omega^{-1} \, \omega_t \, r     \, , \qquad \label{KdV_pre_scs}
\ee
where we introduced $u = 2 \, \phi_x$. From this we obtain the following two versions of a matrix KdV equation with 
self-consistent sources.
\begin{enumerate}
\item Setting $\gamma_1 =0$, i.e., 
\be
      Q^2 \omega = \omega \, P^2 \, ,    \label{KdV_QP_omega_cond}
\ee
and disregarding the equations for $q_t$ and $r_t$, (\ref{KdV_pre_scs}) reduces to 
\be
   4 \, u_t - u_{xxx} - 3 \, (u^2)_x  
     = 8 \, (q \, \omega_t \, r)_x  \, ,  \quad
  q_{xx} = q \, Q^2 - u \, q  \, , \quad
  r_{xx} = P^2 \, r - r \, u \, .    \label{KdV_scs1}
\ee
Here $\omega_t$ can be absorbed by a redefinition of either $q$ or $r$, which exchanges $P$ and $Q$ in the corresponding 
linear equation due to a necessary application of (\ref{KdV_QP_omega_cond}). 
The scalar version appeared in \cite{Mel'88}, also see \cite{Leon+Lati90,LZM01,ZSX03,BFU13}. 
\item $\gamma_2 =0$, i.e., constant $\omega$. In this case (\ref{KdV_pre_scs}) yields
\be
  &&  4 \, u_t - u_{xxx} - 3 \, (u^2)_x  
     = \left( \tilde{q}_x \, r  
       - \tilde{q} \, r_x  \right)_x \, , \nonumber  \\
  && \tilde{q}_t = \tilde{q}_{xxx} + \frac{3}{2} \, u \, \tilde{q}_x  + \frac{3}{4} u_x \, \tilde{q} 
                   + \frac{3}{2} \tilde{q} \, r \, \tilde{q} \, , \quad 
     r_t = r_{xxx} + \frac{3}{2} \, r_x \, u + \frac{3}{4} r \, u_x - \frac{3}{2} r \, \tilde{q} \, r \, ,  \qquad
           \label{KdV_scs2}
\ee
where we introduced $\tilde{q} = 6 \,  q \, (Q^2 \omega - \omega P^2)$. 
This system is also obtained from (\ref{pKP_scs_2_mod}) by assuming that $\phi,q,r$ do not depend on $y$, whereas 
$\omega_y$ is non-zero, but constant. We note that (\ref{KdV_scs2}) is a system of evolution equations for 
all dependent variables. In this respect it is very different from all other examples of systems with 
self-consistent sources presented in this work, with the exception of (\ref{Bouss_scs_2}). 
\end{enumerate}

\paragraph{Exact solutions for vanishing seed.} 
Let $\phi_0$ be constant, i.e., $u_0=0$. Then we have the following solutions of the linear equations,
\be
    \theta = a_1 \, e^{\vartheta(P)} + a_2 \, e^{-\vartheta(P)} \, , \quad
    \eta = e^{\vartheta(Q)} \, b_1 + e^{-\vartheta(Q)} \, b_2 \, , \quad
    \vartheta(P) = P \, x + P^3 \, t \, ,   \label{KdV_theta_eta_sol}
\ee
with constant matrices $a_i,b_i$. A corresponding solution of (\ref{KdV_Omega_eqs}) is 
\bez
   \Omega = e^{\vartheta(Q)} \, A_{11} \, e^{\vartheta(P)} 
            + e^{-\vartheta(Q)} \, A_{22} \, e^{-\vartheta(P)}
            + e^{-\vartheta(Q)} \, A_{21} \, e^{\vartheta(P)}
            + e^{\vartheta(Q)} \, A_{12} \, e^{-\vartheta(P)} + \omega \, ,
\eez
with constant matrices $A_{ij}$ that satisfy the Sylvester equations
\bez
    Q A_{11} + A_{11} P = - b_1 a_1 \, , \quad
    Q A_{22} + A_{22} P = b_2 a_2 \, , \quad
    Q A_{21} - A_{21} P = b_2 a_1 \, , \quad
    Q A_{12} - A_{12} P = -b_1 a_2 \, .
\eez
Then (\ref{phi,q,r}) yields exact solutions of (\ref{KdV_scs1}), respectively (\ref{KdV_scs2}), under the 
corresponding condition for $\omega$. 

\begin{example}
\label{ex:KdV1}
In case of the first version of a KdV equation with self-consistent sources, for the sake of 
definiteness we choose $\omega(t) = \beta(t) \, I_n$. Then (\ref{KdV_QP_omega_cond}) leads 
to $Q = \pm P$. The two cases turn out to be equivalent, hence we choose $Q = P$. Now (\ref{phi,q,r}) 
determines a class of exact solutions of (\ref{KdV_scs1}). Fig.~\ref{fig:KdVscs_2soliton} 
shows an example of a modulated 2-soliton solution. Introducing $\tilde{r} = \beta \, r$, this 
becomes a solution of (\ref{KdV_scs1}) without the presence of $\omega_t$. 

\begin{figure} 
\includegraphics[scale=.25]{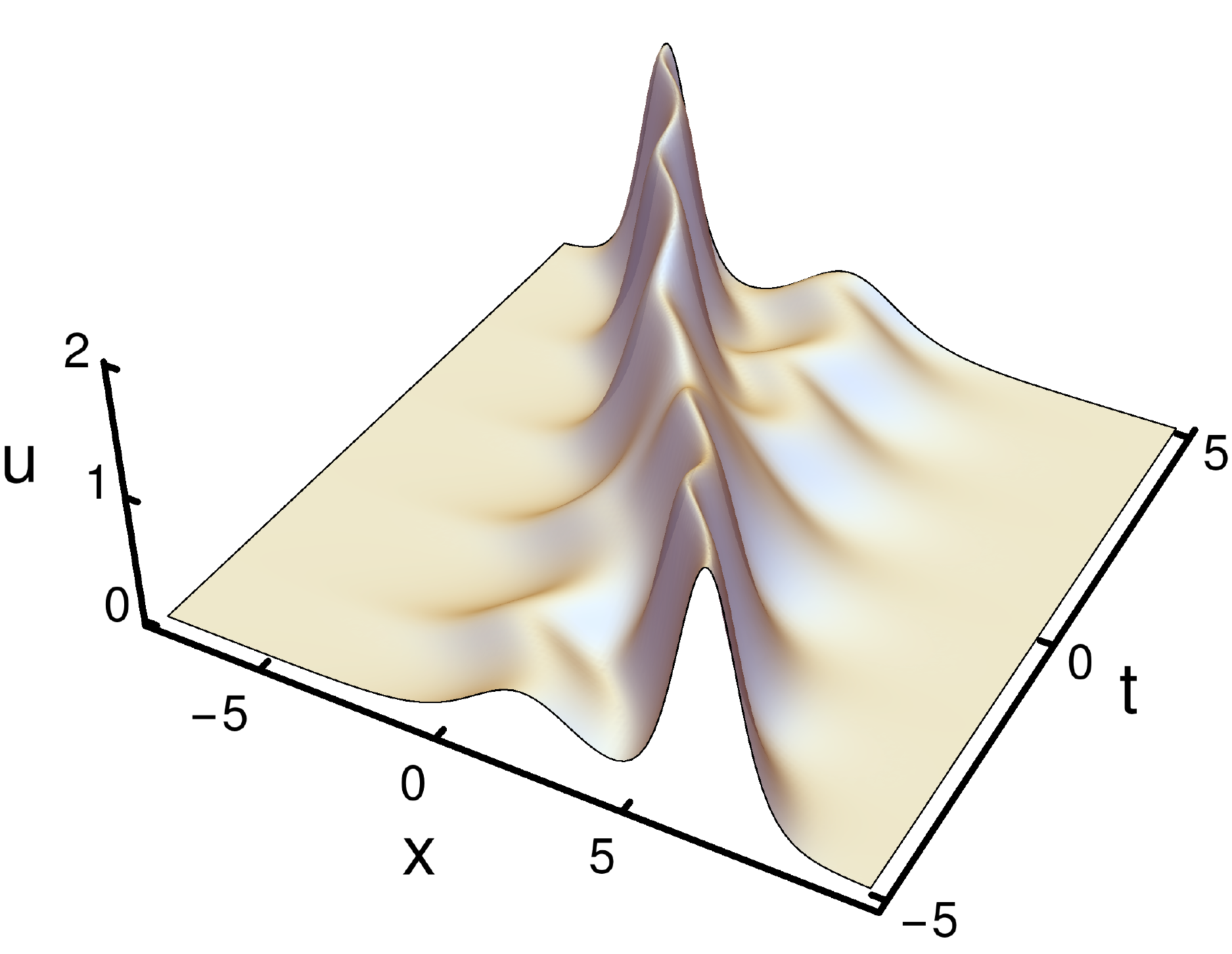} 
\hspace{.2cm}
\includegraphics[scale=.25]{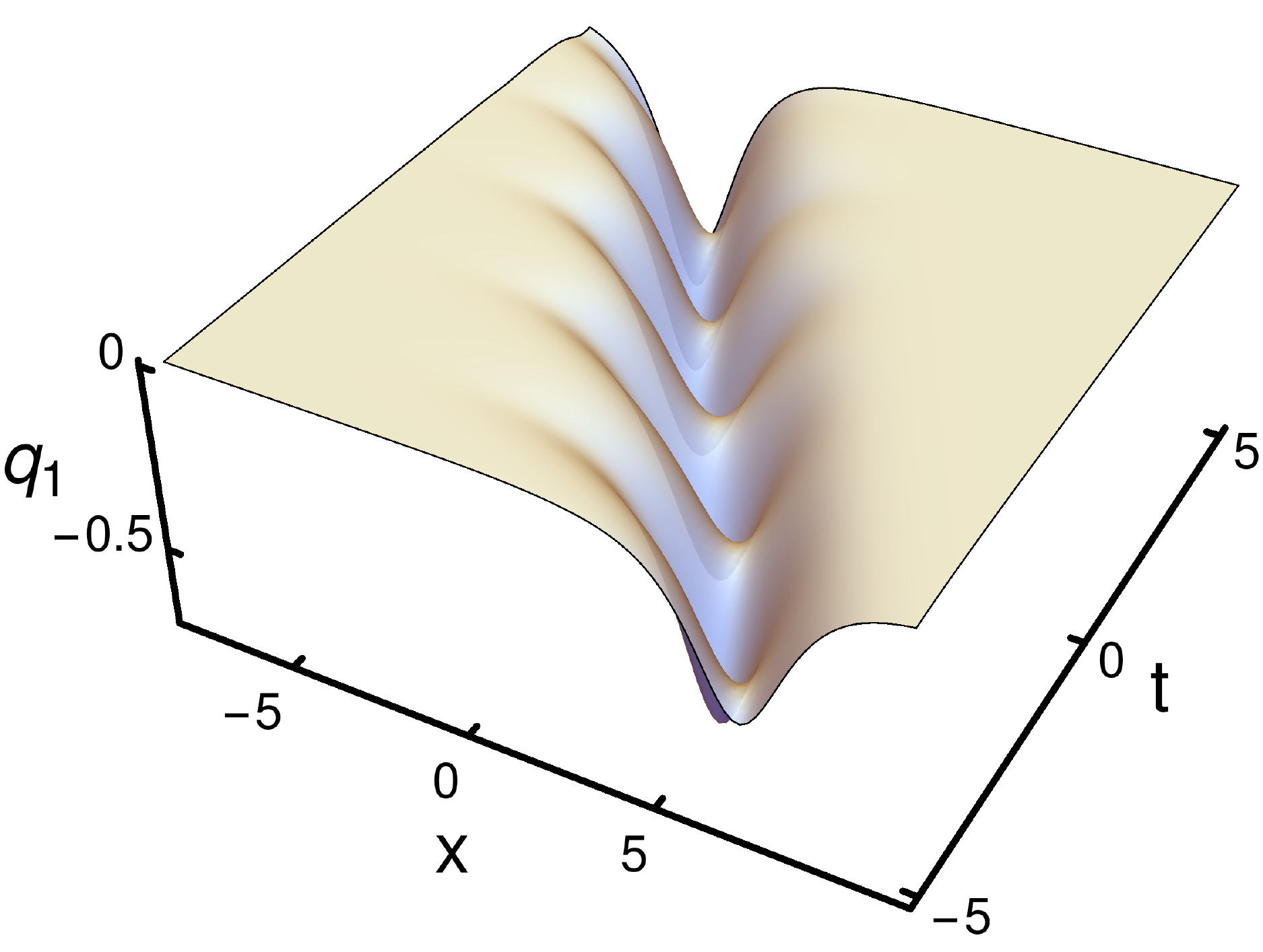} 
\hspace{.2cm}
\includegraphics[scale=.25]{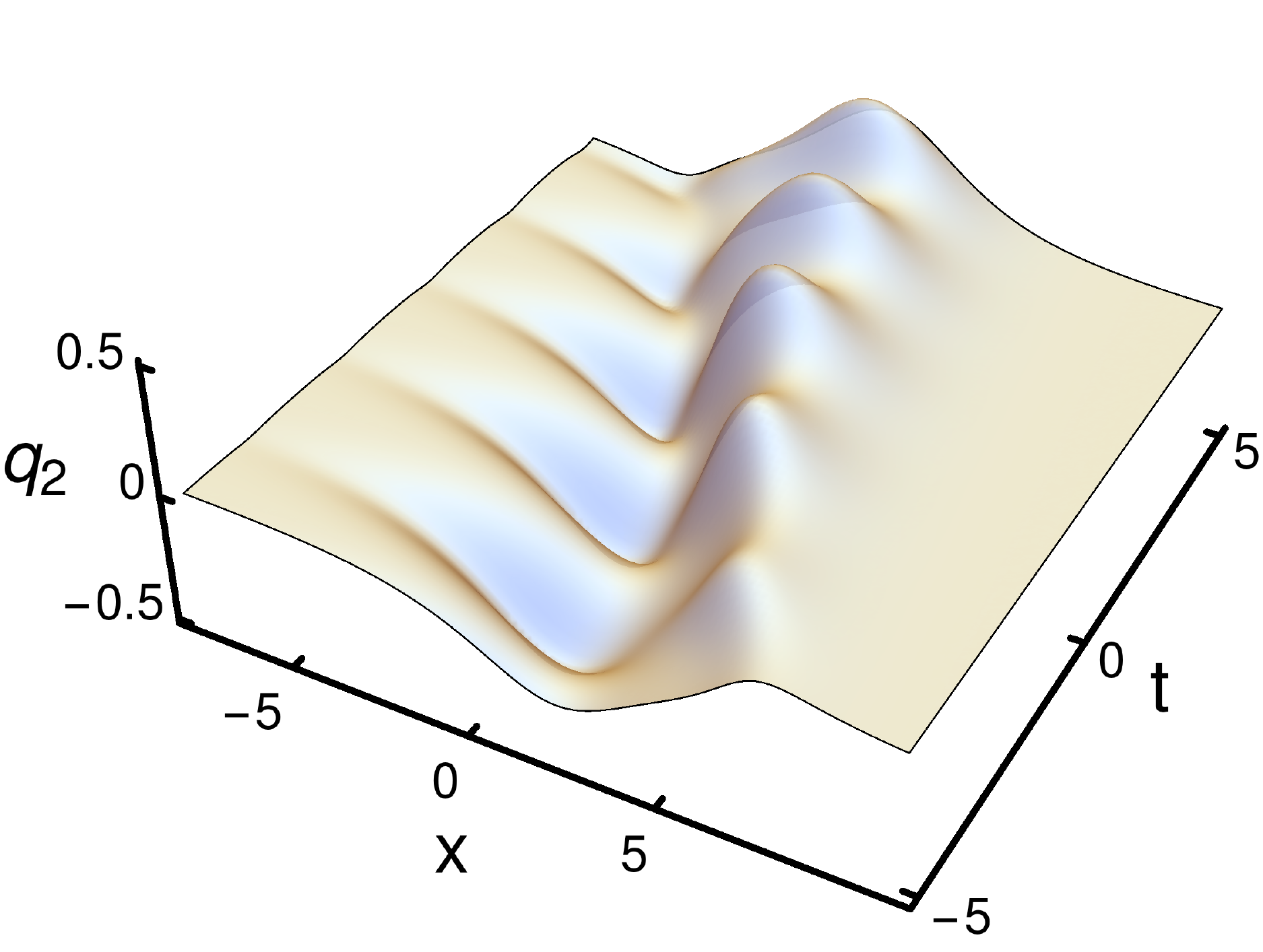} 
\caption{Plots of a 2-soliton solution of the first scalar KdV 
equation with self-consistent sources, see Example~\ref{ex:KdV1}.
Here we chose $\beta(t) = \sin(3 t)$, $p_1=1$, $p_2 = -1/2$, $a_1 = b_1^\intercal = (1,0)$
and $a_2 = (0,1) = b_2^\intercal$ in (\ref{KdV_theta_eta_sol}). 
The plots for $r_1$ and $r_2$ look like those for $q_1$ and $q_2$.
}
\label{fig:KdVscs_2soliton} 
\end{figure}
\end{example}

\begin{example}
In case of the second KdV equation with self-consistent sources, let $n=1$. Then we obtain the following 
2-soliton solution of (\ref{KdV_scs2}),
\bez
  && \phi = \frac{ 2 \, (P^2-Q^2) }{D} \, \cosh(\vartheta_1) \, \sinh(\vartheta_2) \, , \quad
     \tilde{q} = \frac{ e^{-\nu_1 - \nu_2} \, (P+Q) \, (Q^2 \omega - \omega P^2) }{D} \, \cosh(\vartheta_1)  \, , \\ 
  && r =  \frac{e^{-\mu_1 - \mu_2} \, (P-Q) }{D} \, \sinh(\vartheta_2)  \, , \quad
     D := \omega + (Q-P) \, \cosh(\vartheta_1 + \vartheta_2) + (Q+P) \, \cosh(\vartheta_1-\vartheta_2) \, , 
\eez
with constants $\mu_i,\nu_i$ and $\vartheta_1 = \vartheta(P) + \mu_1 - \mu_2$, $\vartheta_2 = \vartheta(Q) + \nu_1 - \nu_2$.
\end{example}

\section{Matrix Boussinesq equation with self-consistent sources}
\label{sec:Bouss}
\setcounter{equation}{0}
Let $\cA_0$ be the space of smooth complex functions on $\mathbb{R}^2$, with coordinates $x$ and $t$, and 
$\cA = \cA_0[\pa]$, where $\pa:=\pa_x$ is the partial differentiation operator 
with respect to $x$. On $\cA$ we define a bidifferential calculus via  
\bez
    \d f = [\pa , f] \, \xi_1 + \frac{1}{2} \, [\pa_t + \pa^2 , f] \, \xi_2 \, , \qquad
    \bd f = \frac{1}{2} \, [\pa_t - \pa^2 , f] \, \xi_1
            - \frac{1}{3} \, [\pa^3,f] \, \xi_2 \, .   
\eez
Setting 
\bez
      \Delta = \Gamma = -I_n \pa \, ,   \quad 
      \kappa = - \frac{1}{3} \, Q^3 \, \xi_2 \, ,   \quad 
      \lambda = -\frac{1}{3} \, P^3 \, \xi_2 \, ,
\eez 
with constant matrices $P,Q$, 
(\ref{Delta,lambda,Gamma,kappa_eqs}) is satisfied, and (\ref{G.o=o.D}) becomes $\omega_x =0$. 
The second equation in (\ref{c,gamma_in_terms_of_omega}) yields
\bez
      \gamma = \gamma_1 \, \xi_1 + (\gamma_2 + \gamma_1 \, \pa) \, \xi_2 \, , \qquad
      \gamma_1 = \frac{1}{2} \, \omega_t \, , \quad
      \gamma_2 = \frac{1}{3} \, (Q^3 \omega + \omega \, P^3) \, .
\eez
The linear equations (\ref{theta_eq}) and (\ref{eta_eq}) read
\bez
  &&  \theta_{xxx} = \theta \, P^3- 3 \, \phi_{0,x} \, \theta_x - \frac{3}{2} (\phi_{0,t} + \phi_{0,xx}) \, \theta   \, , \qquad
    \theta_t = \theta_{xx} + 2 \, \phi_{0,x} \, \theta \, , \\
  && \eta_{xxx} = Q^3 \eta - 3 \, \eta_x \, \phi_{0,x} + \frac{3}{2} \eta \, (\phi_{0,t} - \phi_{0,xx})  \, , \qquad
     \eta_t = -\eta_{xx} - 2 \, \eta \, \phi_{0,x} \, ,
\eez
and (\ref{scs_Omega_eqs}) takes the form
\bez
  && Q^3 (\Omega - \omega) + (\Omega - \omega) \, P^3 = - \eta \, \theta_{xx} + \eta_x \, \theta_x - \eta_{xx} \, \theta 
              - 3 \, \eta \, \phi_{0,x} \, \theta    \, , \nonumber \\
  && \Omega_x = -\eta \, \theta  \, , \qquad
     (\Omega - \omega)_t = - \eta \, \theta_x + \eta_x \, \theta \, .   
\eez
Our general results in Section~\ref{sec:scs_via_bidiff} imply that $\phi$, $q$ and $r$, 
given by (\ref{phi,q,r}), satisfy\footnote{Up to the term involving $Q$ and $P$, the first equation is obtained as a reduction 
of (\ref{pKP_scs}), exchanging $y$ and $t$. } 
\bez
  &&  3 \, \phi_{tt} + \phi_{xxxx} + 6 \, (\phi_x{}^2)_x - 6 \, [\phi_x , \phi_t]  
     = - \left ( 4 \, q \, (Q^3 \omega + \omega \, P^3)\, r + 3 \, q \, \omega_t \, r_x - 3 \, q_x \omega_t \, r \right)_x 
         + 3 \, ( q \, \omega_t \, r)_t \, , \nonumber  \\
  && q_{xxx} = - q \, Q^3 - 3 \, \phi_x q_x - \frac{3}{2} (\phi_t + \phi_{xx}) \, q + \frac{3}{2} q \, \omega_t r \, q
               + q \, (Q^3 \omega + \omega \, P^3) \, \Omega^{-1}  \, , \nonumber \\
  && q_t = q_{xx} + 2 \, \phi_x q - q \, \omega_t \, \Omega^{-1}  \, , \nonumber \\ 
  && r_{xxx} = -P^3 \, r - 3 \, r_x \, \phi_x + \frac{3}{2} r \, (\phi_t-\phi_{xx}) 
               - \frac{3}{2} r \, q \, \omega_t \, r + \Omega^{-1} \, (Q^3 \omega + \omega \, P^3) \, r \, ,   \nonumber \\
  && r_t = - r_{xx} - 2 \, r \, \phi_x - \Omega^{-1} \, \omega_t \, r \, . 
\eez
 From this we obtain the following two versions of a matrix Boussinesq equation with 
self-consistent sources. Here we set $u := 2 \phi_x$. 
\begin{enumerate}
\item $Q^3 \omega + \omega \, P^3 =0$.  
\bez
  &&  3 \, u_{tt} + u_{xxxx} + 3 \, (u^2)_{xx} - 3 \, [u , \pa^{-1} u_t]_x  
     = - 6 \, \left( q \, \omega_t \, r_x - q_x \omega_t \, r \right)_{xx} + 6 \, ( q \, \omega_t \, r)_{xt} \, , \nonumber  \\
  && q_{xxx} = - q \, Q^3 - \frac{3}{2} \, u \, q_x - \frac{3}{4} (\pa^{-1} u_t + u_x) \, q + \frac{3}{2} q \, \omega_t r \, q
               \, , \nonumber\\
  && r_{xxx} = -P^3 \, r - \frac{3}{2} \, r_x \, u + \frac{3}{4} r \, (\pa^{-1}u_t-u_x) 
               - \frac{3}{2} r \, q \, \omega_t r \, .
\eez
The scalar version appeared in \cite{WZF08}. 
\item $\omega_t=0$, i.e., constant $\omega$.
\be
  &&  3 \, u_{tt} + u_{xxxx} + 3 \, (u^2)_{xx} - 3 \, [u , \pa^{-1}u_t]_x   
     = \left( \tilde{q} \, r \right)_{xx} \, , \nonumber  \\
  && \tilde{q}_t = \tilde{q}_{xx} + u \, \tilde{q} \, , \qquad 
     r_t = - r_{xx} - r \, u \, .    \label{Bouss_scs_2}
\ee
where we introduced $\tilde{q} = - 8 \, q \, (Q^3 \omega + \omega \, P^3)$. In contrast to most systems with self-consistent 
sources obtained in this work, here we have evolution equations for \emph{all} dependent variables. Also see (\ref{KdV_scs2}).
\end{enumerate}

\paragraph{Solutions with vanishing seed.} If $u_0 = \phi_{0,x}/2 =0$, solutions of the linear equations are given by 
\bez
  \theta = a_0 \, e^{\vartheta(P)} + a_1 \, e^{\vartheta(\zeta \, P)} + a_2 \, e^{\vartheta(\zeta^2 \, P)} \, , \qquad
  \eta = e^{-\vartheta(-Q)} \, b_0 + e^{-\vartheta(-\zeta \, Q)} \, b_1  + e^{-\vartheta(\zeta^2 \, Q)} \, b_2 \, ,
\eez
where $a_i,b_i$ are constant and 
\bez
    \zeta := e^{2\pi \imag/3} \, , \qquad
    \vartheta(P) := P \, x + P^2 \, t \, .
\eez
A corresponding solution of the equations for $\Omega$ is then given by
\bez
    \Omega = \sum_{j,k=0}^2 e^{-\vartheta(-\zeta^k Q)} \, A_{kj} \, e^{\vartheta(\zeta^j P)} + \omega \, ,
\eez
where $A_{kj}$ are constant matrices subject to
\bez
     \zeta^k Q \, A_{kj} + \zeta^j A_{kj} \, P = - b_k \, a_j  \qquad \quad j,k = 0,1,2 \, .
\eez
Now (\ref{phi,q,r}) yields exact solutions of the above Boussinesq systems with self-consistent sources. 
In case of the first version, we still have to take the constraint $Q^3 \omega = - \omega \, P^3$ into 
account.

\section{Matrix sine-Gordon equation with self-consistent sources}
\label{sec:sG}
Let $\cA_0$ be an associative algebra, where the elements depend on variables $x$ and $y$. Let  
$\cA = \cA_0[J]$, where $J$ is an idempotent operator that determines an involution ${}^\ast$ via 
$J f =: f^\ast \, J$, for $f \in \cA_0$, i.e., $f^\ast = J f J$. 
A bidifferential calculus is then determined on $\cA$ by setting 
\bez
    \d f = [\pa_x,f] \, \xi_1 + \frac{1}{2} [J,f] \, \xi_2 \, , \qquad
    \bd f = \frac{1}{2} [J,f] \, \xi_1 + [\pa_y,f] \, \xi_2 \, .
\eez
In order to eliminate explicit appearances of the operator $J$ in the equations resulting from 
the linear equations (\ref{theta_eq}) and (\ref{eta_eq}), we write\footnote{Only in this section we 
carry the full freedom in $\kappa$ and $\lambda$ along with us, in particular in order to demonstrate that there are 
special non-zero choices, here given by $\kappa_i = \lambda_i =0$, which considerably simplify the equations.}
\bez
  &&  \phi = \varphi \, J \, , \quad
    \Delta = P \, J \, , \quad
    \Gamma = J \, Q \, , \\  
  &&  \kappa = J \, (\kappa_1 + \frac{1}{2} I ) \, \xi_1 + (\kappa_2 - \frac{1}{2} Q^\ast) \, \xi_2 \, , \quad
    \lambda = (\lambda_1 - \frac{1}{2} I ) \, J \, \xi_1 + (\lambda_2 + \frac{1}{2} P^\ast ) \, \xi_2 \, ,
\eez
where $\varphi, P, Q, \kappa_i, \lambda_i$ can now be taken to be matrices over $\cA_0$. We will assume that 
$P$ and $Q$ are invertible. Then we obtain
\be
 &&  \theta_x = \Big( \frac{1}{2} I - \varphi_{0,x} \Big) \, \theta^\ast \, P^{-1} 
                - \theta \, \lambda_1 \, P^{-1}    \, , \quad
     \theta_y = \frac{1}{2} \theta^\ast \, P^\ast - \frac{1}{2} ( \varphi_0 - \varphi_0^\ast ) \, \theta 
                + \theta \, \lambda_2 \, , \nonumber \\
  &&  \eta_x =  Q^{-1} \eta^\ast \Big( \varphi^\ast_{0,x} - \frac{1}{2} I ) - Q^{-1} \kappa_1 \, \eta \, , \quad
      \eta_y = - \frac{1}{2} Q^\ast \eta^\ast  
               + \frac{1}{2} \eta \, ( \varphi_0 - \varphi_0^\ast ) + \kappa_2 \, \eta \, ,  
               \label{sG_theta,eta_eqs}
\ee
where $\theta$ and $\eta$ can now be restricted to be matrices over $\cA_0$. 
The equations (\ref{Delta,lambda,Gamma,kappa_eqs}) impose the following conditions,
\bez
  &&  P_x = ( \lambda_1 P^\ast - P \, \lambda_1^\ast) \, (P^\ast)^{-1} \, , \qquad
      P_y = P \, \lambda_2^\ast - \lambda_2 \, P \, , \nonumber \\
  &&  Q_x = (Q^\ast)^{-1} ( Q^\ast \, \kappa_1 - \kappa_1^\ast \, Q ) \, , \qquad
      Q_y = \kappa_2^\ast \, Q - Q \, \kappa_2 \, , \nonumber \\
  && \kappa_{1,y} + Q \, \kappa_{2,x} = \kappa_2^\ast \, \kappa_1 - \kappa_1 \, \kappa_2  \, , \quad \;\;
      \lambda_{1,y} + \lambda_{2,x} \, P = \lambda_1 \, \lambda_2^\ast - \lambda_2 \, \lambda_1 
       \, . 
\eez
Next we set
\bez
    \Omega = \tilde{\Omega} \, J \, , \quad
    \omega = \tilde{\omega} \, J \, , \quad
    q = \tilde{q} \, J \, , \quad r = J \, \tilde{r}  \, .
\eez
Then (\ref{G.o=o.D}) becomes 
\be
     Q \, \tilde{\omega} = \tilde{\omega}^\ast \, P \, ,   \label{sG_Q,P,om_constr}
\ee
and the involution property implies $Q^\ast Q \, \tilde{\omega} = \tilde{\omega} \, P^\ast P$. 
The second equation in (\ref{c,gamma_in_terms_of_omega}) reads
\bez
      \gamma = \gamma_1 \, \xi_1 + \gamma_2 \, J \, \xi_2 \quad \mbox{where} \quad
      \gamma_1 =  - ( \tilde{\omega}_x + Q^{-1} \kappa_1 \, \tilde{\omega} ) \, P^\ast 
                  - \tilde{\omega} \, \lambda_1^\ast \, , \quad
      \gamma_2 = \tilde{\omega}_y - \tilde{\omega} \, \lambda_2^\ast - \kappa_2 \, \tilde{\omega} \, .
\eez
 From (\ref{scs_Omega_eqs}) we obtain the Sylvester equation
\be
     Q^\ast \, \tilde{\Omega}^\ast - \tilde{\Omega} \, P^\ast = \eta \, \theta \, ,  
     \label{sG_Omega_1}
\ee
and 
\be
 && \tilde{\Omega}_x 
    = Q^{-1} \eta^\ast \, \Big( \frac{1}{2} I - \varphi_{0,x}^\ast \Big) \, \theta \, (P^\ast)^{-1} 
      - \gamma_1 \, (P^\ast)^{-1} - \tilde{\Omega} \, \lambda_1^\ast (P^\ast)^{-1}
      - Q^{-1} \, \kappa_1 \, \tilde{\Omega} \, , \nonumber \\ 
 && \tilde{\Omega}_y = - \frac{1}{2} \, \eta \, \theta^\ast + \gamma_2 
    + \tilde{\Omega} \, \lambda_2^\ast + \kappa_2 \, \tilde{\Omega}  \, .   
    \label{sG_Omega_2}
\ee
Given solutions $\theta$ and $\eta$ of the linear equations, and a solution $\Omega$ of the latter 
consistent system, it follows from our general results in Section~\ref{sec:scs_via_bidiff} that 
\be
     \varphi = \varphi_0 - \theta \, (\tilde{\Omega}^\ast)^{-1} \, \eta^\ast \, , \qquad
     \tilde{q} = \theta \, (\tilde{\Omega}^\ast)^{-1} \, , \qquad
     \tilde{r} = \tilde{\Omega}^{-1} \, \eta    
     \label{sG_scs_sol_varphi}
\ee
solve 
\bez
     \varphi_{xy} - \frac{1}{2} (\varphi - \varphi^\ast) + \frac{1}{2} \, \{ \varphi - \varphi^\ast \, , \, \varphi_x \}
   = \frac{1}{2} (\tilde{q} \, \gamma_1^\ast \, \tilde{r} 
     - \tilde{q}^\ast \, \gamma_1 \, \tilde{r}^\ast)
     + (\tilde{q} \, \gamma_2^\ast \, \tilde{r}^\ast)_x \, ,
\eez
which is obtained from (\ref{scs_phi_eq}), and the system
\bez
     \tilde{q}_x 
 &=& ( \frac{1}{2} I - \varphi_x ) \, \tilde{q}^\ast Q^{-1} 
      + \tilde{q} \, \gamma_1^\ast \, \tilde{\Omega}^{-1} Q^{-1} + \tilde{q} \, (Q^\ast)^{-1} \kappa_1^\ast \, , \\
     \tilde{q}_y 
 &=& \frac{1}{2} \tilde{q}^\ast \, Q^\ast - \frac{1}{2} ( \varphi - \varphi^\ast ) \, \tilde{q} 
      - \tilde{q} \, \gamma_2^\ast \, (\tilde{\Omega}^\ast)^{-1} - \tilde{q} \,   \kappa_2^\ast  \, ,  \\
     \tilde{r}_x 
 &=& P^{-1} \, \tilde{r}^\ast \, (\varphi_x^\ast - \frac{1}{2} I ) 
     + P^{-1} (\tilde{\Omega}^\ast)^{-1} \, \gamma_1^\ast \, \tilde{r} + \lambda_1^\ast (P^\ast)^{-1} \tilde{r} \, , \\
     \tilde{r}_y 
 &=& - \frac{1}{2} P^\ast \, \tilde{r}^\ast + \frac{1}{2} \tilde{r} \, (\varphi - \varphi^\ast)  
     - \tilde{\Omega}^{-1} \, \gamma_2 \, \tilde{r} - \lambda_2^\ast \, \tilde{r} \, ,
\eez
which results from (\ref{scs_q,r_eqs}). $\{ \; , \; \}$ denotes the anti-commutator. 

The extended Miura equation
\bez
     \varphi_x = \frac{1}{2}  (I - g \, (g^\ast)^{-1} ) + \tilde{q} \, \gamma_1^\ast \, P^{-1} \tilde{r}^\ast \, , \qquad
    \frac{1}{2} ( \varphi - \varphi^\ast) = - g_y \, g^{-1} + \tilde{q} \, \gamma_2^\ast \, (P^\ast)^{-1} \, \tilde{r} \, ,
\eez
obtained from (\ref{ext_Miura}), turns the last system into
\bez
      \tilde{q}_x &=& \frac{1}{2} g \, (g^\ast)^{-1} \, \tilde{q}^\ast \, Q^{-1}   
                     - \tilde{q} \, \gamma_1^\ast \, \left( P^{-1} \, \tilde{r}^\ast \, \tilde{q}^\ast
                       - \tilde{\Omega}^{-1} \right)  \, Q^{-1} 
                     + \tilde{q} \, (Q^\ast)^{-1} \kappa_1^\ast  \, , \\
     \tilde{q}_y &=& g_y \, g^{-1} \, \tilde{q} + \frac{1}{2} \tilde{q}^\ast \, Q^\ast 
                     - \tilde{q} \, \gamma_2^\ast \, \left( (P^\ast)^{-1} \, \tilde{r} \, \tilde{q}
                       + (\tilde{\Omega}^\ast)^{-1} \right) - \tilde{q} \, \kappa_2^\ast  \, , \\
     \tilde{r}_x &=& - \frac{1}{2} P^{-1} \tilde{r}^\ast \, g^\ast \, g^{-1} 
                     + P^{-1} \left( \tilde{r}^\ast \, \tilde{q}^\ast \gamma_1 \, (P^\ast)^{-1} 
                         + (\tilde{\Omega}^\ast)^{-1} \gamma_1^\ast  \right) \, \tilde{r} 
                         + \lambda_1^\ast (P^\ast)^{-1} \tilde{r}   \, , \\
     \tilde{r}_y &=& - \tilde{r} \, g_y \, g^{-1} 
                     - \frac{1}{2} P^\ast \, \tilde{r}^\ast
                     + \left( \tilde{r} \, \tilde{q} \, \gamma_2^\ast (P^\ast)^{-1} 
                            - \tilde{\Omega}^{-1} \gamma_2 \right) \, \tilde{r} 
                     - \lambda_2^\ast \, \tilde{r}       \, ,
\eez
and (\ref{scs_g_eq}) has the form 
\be
     (g_y \, g^{-1})_x - \frac{1}{4} \left( g \, (g^\ast)^{-1} - g^\ast \, g^{-1} \right) 
   = \frac{1}{2} \left( \tilde{q} \, \gamma_1^\ast \, P^{-1} \, \tilde{r}^\ast 
                         - \tilde{q}^\ast \, \gamma_1 \, (P^\ast)^{-1} \, \tilde{r} \right)
     + \left( \tilde{q} \, \gamma_2^\ast (P^\ast)^{-1} \, \tilde{r} \right)_x    \, .
                          \label{mSGscs}
\ee
The latter system is solved by 
\be
     g = I - \theta \, (\tilde{\Omega}^\ast)^{-1} (Q^\ast)^{-1} \eta \, g_0 \, , \qquad
     \tilde{q} = \theta \, (\tilde{\Omega}^\ast)^{-1} \, , \qquad
     \tilde{r} = \tilde{\Omega}^{-1} \, \eta \, ,  \label{sG_scs_sol_g}
\ee
if $g_0$ is a solution of the source-free version of (\ref{mSGscs}). 
Equations with self-consistent sources are now obtained as follows.

\begin{enumerate}
\item  $\gamma_2 = 0$. In this case we have
\bez
   &&  \varphi_{xy} - \frac{1}{2} (\varphi - \varphi^\ast) + \frac{1}{2} \, \{ \varphi - \varphi^\ast \, , \, \varphi_x \}
   = \frac{1}{2} (\tilde{q} \, \gamma_1^\ast \, \tilde{r} 
     - \tilde{q}^\ast \, \gamma_1 \, \tilde{r}^\ast)  \, , \\
   && \tilde{q}_y = \frac{1}{2} \tilde{q}^\ast \, Q^\ast - \frac{1}{2} ( \varphi - \varphi^\ast ) \, \tilde{q} 
      - \tilde{q} \, \kappa_2^\ast  \, , \quad
      \tilde{r}_y = - \frac{1}{2} P^\ast \, \tilde{r}^\ast + \frac{1}{2} \tilde{r} \, (\varphi - \varphi^\ast)  
       - \lambda_2^\ast \, \tilde{r} \, , 
\eez
and 
\be
  &&  (g_y \, g^{-1})_x - \frac{1}{4} \left( g \, (g^\ast)^{-1} - g^\ast \, g^{-1} \right) 
   = \frac{1}{2} \left( \tilde{q} \, \gamma_1^\ast \, P^{-1} \, \tilde{r}^\ast 
                         - \tilde{q}^\ast \, \gamma_1 \, (P^\ast)^{-1} \, \tilde{r} \right)
                                 \, , \nonumber \\
  && \tilde{q}_y = g_y \, g^{-1} \, \tilde{q} + \frac{1}{2} \tilde{q}^\ast \, Q^\ast - \tilde{q} \, \kappa_2^\ast  \, ,
          \quad
     \tilde{r}_y = - \tilde{r} \, g_y \, g^{-1} - \frac{1}{2} P^\ast \, \tilde{r}^\ast - \lambda_2^\ast \, \tilde{r}      
         \, .     \label{msG1scs}                         
\ee
\item  $\gamma_1 =0$. Then we have
\bez
   &&  \varphi_{xy} - \frac{1}{2} (\varphi - \varphi^\ast) + \frac{1}{2} \, \{ \varphi - \varphi^\ast \, , \, \varphi_x \}
   = (\tilde{q} \, \gamma_2^\ast \, \tilde{r}^\ast)_x \, , \\
   && \tilde{q}_x = ( \frac{1}{2} I - \varphi_x ) \, \tilde{q}^\ast Q^{-1} + \tilde{q} \, (Q^\ast)^{-1} \kappa_1^\ast
         \, , \quad
      \tilde{r}_x = P^{-1} \, \tilde{r}^\ast \, (\varphi_x^\ast - \frac{1}{2} I ) 
                    + \lambda_1^\ast (P^\ast)^{-1} \tilde{r}     \, ,
\eez 
and
\be
   &&  (g_y \, g^{-1})_x - \frac{1}{4} \left( g \, (g^\ast)^{-1} - g^\ast \, g^{-1} \right) 
   = \left( \tilde{q} \, \gamma_2^\ast (P^\ast)^{-1} \, \tilde{r} \right)_x \, , \nonumber \\
   && \tilde{q}_x = \frac{1}{2} g \, (g^\ast)^{-1} \, \tilde{q}^\ast \, Q^{-1}   
                      + \tilde{q} \, (Q^\ast)^{-1} \kappa_1^\ast  \, , \quad
      \tilde{r}_x = - \frac{1}{2} P^{-1} \tilde{r}^\ast \, g^\ast \, g^{-1} 
                       + \lambda_1^\ast (P^\ast)^{-1} \tilde{r}  \, .  \label{msG2scs}
\ee
\end{enumerate}
To turn this into a concrete system of PDEs, we have to specify $\cA_0$ and the operator $J$. We note that for a \emph{non-zero} 
choice of $\kappa$ and $\lambda$, corresponding to $\kappa_i = \lambda_i=0$, $i=1,2$, the above equations attain a particularly 
simple form.

\paragraph{Scalar sine-Gordon equations with self-consistent sources.}
Let $C^\infty(\bbR^2)$ be the algebra of smooth complex functions of real variables $x$ and $y$. Let  
$\sigma_j$, $j=1,2,3$, be the Pauli matrices, and 
$\cA_0 = \{ a \, I_2 + b \, \sigma_2 \, | \, a,b \in C^\infty(\bbR^2) \}$, which is a commutative algebra. 
We choose $J = \sigma_3$. Note that, for $f \in \cA_0$, also $f^\ast \in \cA_0$. Let
\bez
     g = e^{\imag \, \sigma_2 \, u/2} \, , 
\eez
with a complex function $u$. Then we have $g^\ast = e^{- \imag \, \sigma_2 \, u/2}$ and, for vanishing sources, 
(\ref{mSGscs}) becomes the complex sine-Gordon equation $u_{xy} = \sin u$. 
Any $v \in \cA_0$ has a unique decomposition $v = [v]_0 \, I_2 + [v]_1 \, \imag \, \sigma_2$, and we have 
$v^\ast = [v]_0 \, I_2 - [v]_1 \, \imag \, \sigma_2$.  

\begin{enumerate}
\item $\gamma_2 = 0$. Setting $\kappa_2 = \lambda_2=0$, from (\ref{msG1scs}) we obtain
\bez
  &&  u_{xy} - \sin u = - 2 \, [ \tilde{q} \, \tilde{\omega}_x^\ast \, \tilde{r}^\ast ]_1 \, , \\
  &&  \tilde{q}_y = \frac{\imag}{2} \, u_y \, \sigma_2 \, \tilde{q} + \frac{1}{2} \tilde{q}^\ast \, Q^\ast \, , \quad
    \tilde{r}_y = - \frac{\imag}{2} u_y \, \tilde{r} \, \sigma_2 - \frac{1}{2} P^\ast \, \tilde{r}^\ast  \, . 
\eez
Then $P_y = Q_y =0$, so that $P$ and $Q$ are only allowed to depend on $x$. 
With the decomposition $\tilde{q} = f_1 \otimes I_2 + f_2 \otimes \imag \, \sigma_2$, 
$\tilde{r} = h_2 \otimes I_2 + h_1 \otimes \imag \, \sigma_2$, with row vectors $f_i$ and column vectors $h_i$,   
and assuming $Q^\ast = Q$ and $P^\ast = P$, so that $Q = \tilde{Q} \otimes I_2$ and $P = \tilde{P} \otimes I_2$,
the linear equations for $\tilde{q}$ and $\tilde{r}$ can be written as
\bez
   \left( \begin{array}{cc} \pa_y & u_y/2 \\ u_y/2 & -\pa_y \end{array}\right) 
   \left(\begin{array}{c} f_1 \\ f_2 \end{array}\right)
   = \frac{1}{2} \left(\begin{array}{c} f_1 \, \tilde{Q} \\ f_2 \, \tilde{Q} \end{array}\right)  \, , \quad
   \left( \begin{array}{cc} \pa_y & u_y/2 \\ u_y/2 & -\pa_y \end{array}\right) 
   \left(\begin{array}{c} h_1 \\ h_2 \end{array}\right)
   = \frac{1}{2} \left(\begin{array}{c} \tilde{P} \, f_1 \\ \tilde{P} \, f_2 \end{array}\right)  \, .
\eez
Of course, the constraint (\ref{sG_Q,P,om_constr}) has to be taken into account. Due to the latter, $\tilde{\omega}_x$ 
can be absorbed by a redefinition of either $\tilde{q}$ or $\tilde{r}$, while only exchanging $P$ and $Q$ in the 
respective linear equation. Then we obtain 
\bez
      u_{xy} - \sin u = f_1 h_1 - f_2 h_2 \, ,
\eez      
which establishes contact with the sine-Gordon equation with self-consistent sources considered, e.g., in \cite{Khas+Uraz09}. 
\item $\gamma_1 =0$. Setting $\kappa_1 = \lambda_1 =0$, (\ref{msG2scs}) becomes 
\bez
   && u_{xy} - \sin u = 2 \, \left[ \left( \tilde{q} \, \tilde{\omega}_y^\ast (P^\ast)^{-1} \, \tilde{r} \right)_x \right]_1 \, , \\
   && \tilde{q}_x = \frac{1}{2} e^{\imag \, \sigma_2 \, u} \, \tilde{q}^\ast \, Q^{-1}  \, , \quad 
      \tilde{r}_x = - \frac{1}{2} P^{-1} \tilde{r}^\ast \, e^{- \imag \, \sigma_2 \, u}  \, . 
\eez
In terms of the new variables $\hat{q} = \tilde{q} \, e^{-\imag \, \sigma_2 \, u/2}$ and 
$\hat{r} = e^{\imag \, \sigma_2 \, u/2} \, \tilde{r}$, this becomes
\bez
   &&  u_{xy} - \sin u = 2 \, \left[ ( \hat{q} \, \tilde{\omega}_y^\ast (P^\ast)^{-1} \hat{r} )_x \right]_1 \, , \\
   && \hat{q}_x =  \frac{1}{2} \, \hat{q}^\ast \, Q^{-1} 
                  - \frac{\imag}{2} \, u_x \, \sigma_2 \, \hat{q} \, , \quad
      \hat{r}_x = - \frac{1}{2} P^{-1} \hat{r}^\ast + \frac{\imag}{2} \, u_x \, \sigma_2 \, \hat{r}  \, .
\eez
Then $P$ and $Q$ can only depend on $y$. We can absorb $\tilde{\omega}_y P^{-1}$ by 
a redefinition of $\hat{q}$, but in the linear equation for $\hat{q}$ we have to replace $Q$ by $P^\ast$ (via 
an application of (\ref{sG_Q,P,om_constr})). 
Then, writing $\hat{q} = f_1 \otimes I_2  + f_2 \otimes \imag \, \sigma_2$,  
$\hat{r} = f_2^\intercal \otimes I_2 + f_1^\intercal \otimes \imag \, \sigma_2$ (where $^\intercal$ denotes the 
transpose), and setting $P = (P^\ast)^\intercal$, we obtain 
\bez
     u_{xy} - \sin u = 2 \, ( f_1 f_1^\intercal + f_2 f_2^\intercal )_x \, ,
\eez
and thus the sine-Gordon equation with self-consistent sources considered in \cite{Zhan+Chen03,Zhang03}.
Of course, we still have to respect the constraint (\ref{sG_Q,P,om_constr}). 
\end{enumerate}

With the above choice for $\cA_0$ and $J$, the systems (\ref{msG1scs}) and (\ref{msG2scs}) can thus be regarded 
as matrix generalizations of the above sine-Gordon equations with self-consistent sources. The corresponding systems 
involving $\varphi$ instead of $g$ are Miura-duals of them.

\begin{remark} A simpler choice for $\cA_0$ and the involution ${}^\ast$ is the algebra of smooth complex functions 
on $\mathbb{R}^2$ together with complex conjugation. 
In the scalar case ($m=1$), we then set $g = e^{\imag \, u/2}$, with real $u$. 
This leads to similar results, but there is a restriction to self-consistent source 
extensions of the \emph{real} sine-Gordon equation. The slightly more complicated setting we chose above is more flexible.
\end{remark}

\paragraph{Exact solutions in case of trivial seed.} 
We set $\kappa_1 = \kappa_2 = \lambda_1 = \lambda_2 =0$, and $P = I_2 \otimes \tilde{P}$,  
$Q = \tilde{Q} \otimes I_2$, with constant complex matrices $\tilde{P}$ and $\tilde{Q}$. 
Then we have $P=P^\ast$ and $Q=Q^\ast$, and special solutions of the linear equations (\ref{sG_theta,eta_eqs}) 
with $\varphi_0 =0$ are given by 
\bez
 && \theta = a_0 \, e^{\vartheta(\tilde{P})/2} \otimes I_2 + a_1 \, e^{-\vartheta(\tilde{P})/2} \otimes \imag \, \sigma_2 \, \, , \qquad
 \eta = e^{-\vartheta(\tilde{Q})/2} \, b_0 \otimes I_2 + e^{\vartheta(\tilde{Q})/2} \, b_1 \otimes \imag \, \sigma_2 \, , \\
 &&     \vartheta(K) := x \, K^{-1} + y \, K \, , 
\eez
with constant complex $1 \times n$ matrices $a_i$ and $n \times 1$ matrices $b_i$. 
The equations (\ref{sG_Omega_1}) and (\ref{sG_Omega_2}) are solved by 
$\tilde{\Omega} = [\tilde{\Omega}]_0 \otimes I_2 + [\tilde{\Omega}]_1 \otimes \imag \, \sigma_2$ with 
\bez
  && [\tilde{\Omega}]_0 = e^{-\vartheta(\tilde{Q})/2} \, A_0 \, e^{\vartheta(\tilde{P})/2}
                        + e^{\vartheta(\tilde{Q})/2} \, B_0 \, e^{-\vartheta(\tilde{P})/2} + [\tilde{\omega}]_0 \, , \\
  && [\tilde{\Omega}]_1 = e^{-\vartheta(\tilde{Q})/2} \, A_1 \, e^{-\vartheta(\tilde{P})/2}
                        + e^{\vartheta(\tilde{Q})/2} \, B_1 \, e^{\vartheta(\tilde{P})/2} + [\tilde{\omega}]_1 \, ,                     
\eez
where the constant complex matrices $A_0,A_1,B_0,B_1$ are subject to the Sylvester equations
\bez
     \tilde{Q} A_0 - A_0 \tilde{P} = b_0 \, a_0 \, , \quad
     \tilde{Q} B_0 - B_0 \tilde{P} = b_1 \, a_1 \, , \quad
     \tilde{Q} A_1 + A_1 \tilde{P} = - b_0 \, a_1 \, , \quad
     \tilde{Q} B_1 + B_1 \tilde{P} = - a_0 \, b_1 \, ,
\eez
and $\omega$ has to satisfy (\ref{sG_Q,P,om_constr}). 
Now (\ref{sG_scs_sol_g}), with constant $g_0$, leads to a class of 
exact solutions of the above scalar sine-Gordon equations with self-consistent sources.

\section{Matrix Nonlinear Schr\"odinger equation with self-consistent sources}
\label{sec:NLS}
\setcounter{equation}{0}
Let $\cA$ be the algebra of $M \times M$ matrices of smooth functions of coordinates $x$ and $t$ on 
$\mathbb{R}^2$. We define
a bidifferential calculus on $\cA$ via 
\bez
    \d f = - \frac{1}{2} \, [J , f] \, \xi_1 + [\pa , f] \, \xi_2 \, , \qquad
    \bd f = [\pa , f] \, \xi_1 - [\imag \, \pa_t , f] \, \xi_2 \, ,
\eez 
where $J \neq I$ is a constant $M \times M$ matrix, i.e., $J_x = J_t = 0$,  
satisfying $J^2 = I$ (also see \cite{DMH10NLS}). $\d$ and $\bd$ extend to matrices over $\cA$ and to the 
corresponding graded algebra. We set $m=1$ and 
\bez
  \Delta = P \, , \quad     \Gamma = Q \, ,  \quad
  \kappa = (\kappa_1 - \frac{1}{2} Q J ) \, \xi_1 + (\kappa_2 - Q \, \kappa_1) \, \xi_2 \, , \quad
  \lambda = (\lambda_1 + \frac{1}{2} J P) \, \xi_1 + (\lambda_2 - \lambda_1 P) \, \xi_2 \, ,
\eez
where $P, Q, \kappa_i, \lambda_i$ are matrices over $\cA_0$. 
Then the equations (\ref{Delta,lambda,Gamma,kappa_eqs}) lead to
\bez
  &&  P_x = [P , \lambda_1] \, , \qquad
    \imag \, P_t = - [P , \lambda_2 ] \, , \qquad
    \imag \, \lambda_{1,t} + \lambda_{2,x} = - [\lambda_1 , \lambda_2] \, , \\
  && Q_x = - [Q, \kappa_1] \, , \qquad 
     \imag \, Q_t = [Q , \kappa_2] \, , \qquad
     \imag \, \kappa_{1,t} + \kappa_{2,x} = [\lambda_1 , \lambda_2] \, ,
\eez
(\ref{G.o=o.D}) becomes
\be
      Q \, \omega = \omega P \, ,    \label{NLS_Q,P,omega_cond}
\ee
and the second equation in (\ref{c,gamma_in_terms_of_omega}) reads
\bez
    \gamma = \gamma_1 \, \xi_1 + (\gamma_2 - \gamma_1 P) \, \xi_2 \, , \quad
    \gamma_1 = \omega_x - \kappa_1 \, \omega - \omega \lambda_1 \, , \quad
    \gamma_2 = - \imag \, \omega_t - \kappa_2 \, \omega - \omega \lambda_2 \, .
\eez
The linear equations (\ref{theta_eq}) and (\ref{eta_eq}) take the form
\be
  &&  \theta_x = \frac{1}{2} J \theta P + \theta \lambda_1 + \frac{1}{2} [J,\phi_0] \, \theta \, , \quad
    \imag \, \theta_t = - \frac{1}{2} J \theta P^2 - \theta \lambda_2 - \frac{1}{2} [J,\phi_0] \, \theta P - \phi_{0,x} \, \theta \, , 
           \nonumber \\
  &&  \eta_x = - \frac{1}{2} Q \eta J + \kappa_1 \eta - \frac{1}{2} \eta \, [J , \phi_0] \, , \quad
    \imag \, \eta_t = \frac{1}{2} Q^2 \eta J - \kappa_2 \, \eta + \frac{1}{2} Q \eta \, [J , \phi_0] + \eta \, \phi_{0,x} \, ,
                    \label{NLS_lin_eqs}
\ee
and (\ref{scs_Omega_eqs}) results in
\be
 && Q \Omega - \Omega P = \eta \, \theta \, , \qquad
  \Omega_x = - \frac{1}{2} \eta \, J \, \theta + \kappa_1 \omega + \Omega \lambda_1 + \gamma_1 \, ,  \nonumber \\
 && \imag \, \Omega_t = \frac{1}{2} ( Q \eta \, J \, \theta + \eta J \theta P) - \kappa_2 \, \Omega - \Omega \lambda_2 
                        +\frac{1}{2} \eta \, [J , \phi_0 ] \, \theta - \gamma_2 \, .    \label{NLS_Omega_eqs}
\ee
Now $\phi, q,r$ given by (\ref{phi,q,r}) solve
\bez
    \frac{\imag}{2} [J , \phi_t] + \phi_{xx} - \frac{1}{2} [[J,\phi],\phi_x] 
  = (q \gamma_1 r)_x - \frac{1}{2} [J, q \, (\gamma_2 - \gamma_1 P) \, r ]
\eez
and
\bez
    q_x &=& \frac{1}{2} J q \, Q - q \, \kappa_1 + \frac{1}{2} [J,\phi] \, q - q \, \gamma_1 \, \Omega^{-1} \, , \\
    \imag \, q_t &=& - \frac{1}{2} J q \, Q^2 + q \, \kappa_2 - \phi_x \, q - \frac{1}{2} [J,\phi] \, q \, Q + q \, \gamma_1 \, r \, q
                     + q \gamma_2 \, \Omega^{-1} \, , \\
    r_x &=& - \frac{1}{2} P r J - \lambda_1 r - \frac{1}{2} r \, [J,\phi] - \Omega^{-1} \gamma_1 \, r \, , \\
    \imag \, r_t &=& \frac{1}{2} P^2 r J + \lambda_2 \, r + r \, \phi_x + \frac{1}{2} P r \, [J,\phi] - r \, q \, \gamma_1 \, r 
                     + \Omega^{-1} \gamma_2 \, r \, .
\eez
Next we choose $J = \mbox{block-diag}(I_{M_1} , -I_{M_2} )$, where $M_1+M_2=M$, and use the block decompositions
\be
    \phi = J \left(\begin{array}{cc} u & \mathfrak{q} \\ \mathfrak{r} & v \end{array}\right) 
         = \left(\begin{array}{rr} u & \mathfrak{q} \\ - \mathfrak{r} & - v \end{array}\right) \, , \quad
    q = \left(\begin{array}{c} q_1 \\ q_2 \end{array}\right) \, , \quad 
    r = \left(\begin{array}{cc} r_1 & r_2 \end{array}\right) \, ,  \label{NLS_block-decomp}
\ee
where $u,v,\mathfrak{q},\mathfrak{r}$ are, respectively, $M_1 \times M_1$, $M_2 \times M_2$, 
$M_1 \times M_2$ and $M_2 \times M_1$ matrices. $q_1$ and $q_2$ have size $M_1 \times (M \cdot n)$ and
$M_2 \times (M \cdot n)$, respectively. Then we obtain the following AKNS equations with self-consistent sources. 
Here and in the following we set $\kappa_i = \lambda_i =0$, in which case $P$ and $Q$ have to be constant. 
\begin{enumerate}
\item $\gamma_1 = \omega_x =0$. 
\be
   && \imag \, \mathfrak{q}_t + \mathfrak{q}_{xx} - 2 \, \mathfrak{q} \, \mathfrak{r} \, \mathfrak{q} = \imag \, q_1 \, \omega_t \, r_2 \, , \qquad
      - \imag \, \mathfrak{r}_t + \mathfrak{r}_{xx} - 2 \, \mathfrak{r} \, \mathfrak{q} \, \mathfrak{r} = \imag \, q_2 \, \omega_t \, r_1 \, , 
        \nonumber \\
   && q_{1,x} = \frac{1}{2} \, q_1 \, Q + \mathfrak{q} \, q_2 \, , \quad 
      q_{2,x} = - \frac{1}{2} \, q_2 \, Q + \mathfrak{r} \, q_1 \, ,  \nonumber \\
   && r_{1,x} = - \frac{1}{2} \, P \, r_1 - r_2 \, \mathfrak{r} \, , \quad 
      r_{2,x} = \frac{1}{2} \, P \, r_2 - r_1 \, \mathfrak{q} \, ,   \label{AKNS_scs1}
\ee
and $u = - \mathfrak{q} \, \mathfrak{r}$, $v = - \mathfrak{r} \, \mathfrak{q}$.
\item $\gamma_2 = - \imag \, \omega_t =0$.
\be
  && \imag \, \mathfrak{q}_t + \mathfrak{q}_{xx} - 2 \, \mathfrak{q} \, \mathfrak{r} \, \mathfrak{q} 
      = \mathfrak{q} \, (q_2 \, \omega_x \, r_2) - (q_1 \, \omega_x \, r_1) \, \mathfrak{q} + (q_1 \, \omega_x \, r_2)_x + q_1 \, \omega_x \, P \, r_2 \, , 
               \nonumber \\
  && -\imag \, \mathfrak{r}_t + \mathfrak{r}_{xx} - 2 \, \mathfrak{r} \, \mathfrak{q} \, \mathfrak{r} 
      = -\mathfrak{r} \, (q_1 \, \omega_x \, r_1) + (q_2 \, \omega_x \, r_2) \, \mathfrak{r} - (q_2 \, \omega_x \, r_1)_x + q_2 \, \omega_x \, P \, r_1 \, , 
               \nonumber  \\
  && \imag \, q_{1,t} + \frac{1}{2} \, q_1 \, Q^2 - \mathfrak{q} \, \mathfrak{r} \, q_1 + \mathfrak{q}_x \, q_2 + \mathfrak{q} \, q_2 \, Q 
     = q_1 \, \omega_x \, r_2 \, q_2 \, , \nonumber \\
  && \imag \, q_{2,t} - \frac{1}{2} \, q_2 \, Q^2 + \mathfrak{r} \, \mathfrak{q} \, q_2 - \mathfrak{r}_x \, q_1 + \mathfrak{r} \, q_1 \, Q 
     = q_2 \, \omega_x \, r_1 \, q_1 \, , \nonumber \\
  && \imag \, r_{1,t} - \frac{1}{2} \, P^2 \, r_1 + r_1 \, \mathfrak{q} \, \mathfrak{r} + r_2 \, \mathfrak{r}_x - P \, r_2 \, \mathfrak{r}  
     = - r_2 \, q_2 \, \omega_x \, r_1 \, , \nonumber  \\
  && \imag \, r_{2,t} + \frac{1}{2} \, P^2 \, r_2 - r_2 \, \mathfrak{r} \, \mathfrak{q} - r_1 \, \mathfrak{q}_x - P \, r_1 \, \mathfrak{q}  
     = - r_1 \, q_1 \, \omega_x \, r_2 \, ,   \label{AKNS_scs2}
\ee
and $u_x = - \mathfrak{q} \, \mathfrak{r} + q_1 \, \omega_x \, r_1$, $v_x = - \mathfrak{r} \, \mathfrak{q} - q_2 \, \omega_x \, r_2$.
\end{enumerate}
In both cases, $Q,P$ and $\omega$ still have to satisfy (\ref{NLS_Q,P,omega_cond}). 
Via a Hermitian conjugation reduction, see below, these systems become matrix versions of two kinds of (scalar) Nonlinear Schr\"odinger (NLS) equations  
with self-consistent sources. The second system seems to be new, even in the scalar case.

\paragraph{Hermitian conjugation reductions.}
Imposing the conditions
\be
   u^\dagger = u \, , \quad 
   v^\dagger = v \, , \quad
   \mathfrak{r} = \varepsilon \, \mathfrak{q}^\dagger \, , \quad 
   r_j = \epsilon_j \, q_j^\dagger \qquad j=1,2 \, , \quad
   \varepsilon \, \epsilon_1 \epsilon_2 = -1 \, , \quad
   \omega^\dagger = \omega      \, ,     \label{NLS_decomp_hc_reduction}
\ee
where $\varepsilon, \epsilon_1, \epsilon_2 \in \{-1,1\}$,  and 
\bez
       P = - Q^\dagger \, , 
\eez
we find that (\ref{AKNS_scs1}) and (\ref{AKNS_scs2}) reduce to the following systems.
\begin{enumerate}
\item $\omega_x =0$. 
\be
  \imag \, \mathfrak{q}_t + \mathfrak{q}_{xx} 
       - 2 \varepsilon \, \mathfrak{q} \, \mathfrak{q}^\dagger \, \mathfrak{q} 
     = \imag \, \epsilon_2 \, q_1 \, \omega_t \, q_2^\dagger \, , \quad
     q_{1,x} = \frac{1}{2} \, q_1 \, Q + \mathfrak{q} \, q_2 \, , \quad 
     q_{2,x} = - \frac{1}{2} \, q_2 \, Q + \varepsilon \, \mathfrak{q}^\dagger \, q_1 \, . \label{NLSscs1}
\ee
We can set $\epsilon_2 = 1$, since it can be absorbed into $\omega$. 
In the scalar case, such equations appeared previously in \cite{Dokt+Vlas83,NYK91,CLL91,Mel'91,Mel'92,Shao+Zeng05}.
\item $\omega_t =0$. 
\be
  && \imag \, \mathfrak{q}_t + \mathfrak{q}_{xx} 
        - 2 \varepsilon \, \mathfrak{q} \, \mathfrak{q}^\dagger \, \mathfrak{q} 
      = - \varepsilon \, \epsilon_1 \, \mathfrak{q} \, (q_2 \, \omega_x \, q_2^\dagger) 
        - \epsilon_1 \, (q_1 \, \omega_x \, q_1^\dagger) \, \mathfrak{q} 
        - \varepsilon \, \epsilon_1 \, (q_1 \, \omega_x \, q_2^\dagger)_x 
        + \varepsilon \, \epsilon_1 \, q_1 \, \omega_x \, Q^\dagger \, q_2^\dagger \, , 
               \nonumber \\
  && \imag \, q_{1,t} + \frac{1}{2} \, q_1 \, Q^2 
       - \varepsilon \, \mathfrak{q} \, \mathfrak{q}^\dagger \, q_1 + \mathfrak{q}_x \, q_2 
       + \mathfrak{q} \, q_2 \, Q \,  
     = - \varepsilon \, \epsilon_1 \, q_1 \, \omega_x \, q_2^\dagger \, q_2 \, , \nonumber \\
  && \imag \, q_{2,t} - \frac{1}{2} \, q_2 \, Q^2 + \varepsilon \, \mathfrak{q}^\dagger \, \mathfrak{q} \, q_2 
       - \varepsilon \, \mathfrak{q}_x^\dagger \, q_1 + \varepsilon \, \mathfrak{q}^\dagger \, q_1 \, Q 
     = \epsilon_1 \, q_2 \, \omega_x \, q_1^\dagger \, q_1 \, .  \label{NLSscs2}
\ee 
We can set $\epsilon_1 = 1$, since it can be absorbed into $\omega$. Though $\omega_x$ can be absorbed (e.g., by  
redefinitions of $r_1$ and $r_2$, before the reduction), we cannot simultaneously absorb $Q$ on the right hand 
side of the first equation.
The last system is therefore of a different nature than the familiar integrable equations with self-consistent sources. 
In contrast to (\ref{NLSscs1}), the equations for $q_1$ and $q_2$ are \emph{non}linear in the system  
(\ref{NLSscs2}). 
\end{enumerate}

In both cases, $Q$ and $\omega$ also have to satisfy $Q \, \omega + \omega \, Q^\dagger = 0$, as a consequence of 
(\ref{NLS_Q,P,omega_cond}). This severely restricts $Q$ if $\omega \neq 0$. If $\omega = \beta \, I_{nM}$, 
with a scalar $\beta$, then $Q^\dagger = - Q$. For diagonal $Q$, this restricts its eigenvalues to be imaginary, 
a restriction that also appears in \cite{Shao+Zeng05}, in the focusing NLS case $\varepsilon = -1$. 
$\varepsilon = 1$ is the defocusing case.

The above reduction conditions (\ref{NLS_decomp_hc_reduction}) for $\mathfrak{q},  \mathfrak{r}, u, v, q_j, r_j$ 
can be expressed as follows,
\be
     \phi^\dagger = \Lambda \, \phi \, \Lambda \, , \quad
     r = \epsilon_1 \, q^\dagger \Lambda \, , \quad 
     \Lambda := \left\{ \begin{array}{l} J \\ I \end{array} \mbox{ if } \begin{array}{l} \varepsilon = 1 \\ \varepsilon = -1 \end{array} 
                \right. \, .   \label{NLS_hc_reduction}
\ee
For the solution generating procedure to respect the reduction conditions, we still have to require 
\bez
      \eta = \epsilon_1 \, \theta^\dagger \Lambda \, .
\eez

\paragraph{Exact solutions for vanishing seed.} 
Let $\phi_0 = 0$. The linear equations (\ref{NLS_lin_eqs}) are then solved by
\bez
   \theta = a_+ \, e^{\vartheta(-Q^\dagger)/2} + a_- \, e^{-\vartheta(-Q^\dagger)/2} \, , \quad 
   \eta = e^{-\vartheta(Q)/2} \, b_+ + e^{\vartheta(Q)/2} \, b_- \, , \quad
   \vartheta(Q) = x \, Q J + \imag \, t \, J \, Q^2 \, , 
\eez
where 
\bez
      J \, a_{\pm} = \pm a_{\pm} \, , \quad
      b_+ = \epsilon_1 \, a_+^\dagger \, , \quad
      b_- = - \varepsilon \, \epsilon_1 \, a_-^\dagger  \, .
\eez      
A corresponding solution of (\ref{NLS_Omega_eqs}) is
\bez
    \Omega = e^{-\vartheta(Q)/2} \, A \, e^{\vartheta(-Q^\dagger)/2} + e^{\vartheta(Q)/2} \, B \, e^{-\vartheta(-Q^\dagger)/2}
             + \omega \, ,
\eez
where the constant matrices $A, B$ have to satisfy the Sylvester equations
\bez
   Q \, A + A \, Q^\dagger = \epsilon_1 \, a_+^\dagger a_+ \, , \qquad 
  &&   Q \, B + B \, Q^\dagger = - \varepsilon \epsilon_1 \, a_-^\dagger a_- \, .
\eez
If $Q$ and $-Q^\dagger$ have no eigenvalue in common, the solutions $A$ and $B$ are unique and Hermitian. 
But $Q \omega + \omega Q^\dagger =0$ then implies $\omega=0$. Thus, in order to obtain solutions of (\ref{NLSscs1})
or (\ref{NLSscs2}) with $\omega \neq 0$, the spectrum condition for $Q$ needs to be violated. Requiring  
$A^\dagger = A$ and $B^\dagger = B$, then $\Omega$ is Hermitian, since $\vartheta(Q)^\dagger = - \vartheta(-Q^\dagger)$, 
and $\phi$ satisfies the reduction condition. 
If $\omega$ and $Q$ satisfy $Q \omega + \omega Q^\dagger =0$, and if $\omega_x =0$ or $\omega_t =0$, then   
\bez
     \phi = - \epsilon_1 \, \theta \, \Omega^{-1} \theta^\dagger \Lambda \, , \qquad
     q = \theta \, \Omega^{-1} 
\eez
provides us with exact solutions of the above matrix NLS equations with self-consistent sources. 
In the defocusing NLS case, it is more interesting to start with a constant density seed solution (see, e.g., \cite{Shao+Zeng05}).

\section{Matrix Davey-Stewartson equation with self-consistent sources}
\label{sec:DS}
\setcounter{equation}{0}
Let $\cA_0$ be the algebra of $M \times M$ matrices of smooth complex functions on 
$\mathbb{R}^3$, and $\cA = \cA_0[\pa]$, where 
$\pa$ is again the partial differentiation operator with respect to $x$. We define
a bidifferential calculus on $\cA$ via
\be
    \d f = \frac{1}{2} \, [J , f] \, \xi_1 + [\pa_y + J \pa , f] \, \xi_2 \, , \qquad
    \bd f = [\pa_y - J \pa , f] \, \xi_1 - [\imag \, \pa_t + 2 J \pa^2 , f] \, \xi_2 \, ,
    \label{DS_bidiff}
\ee 
where $J \neq I$ is a constant $M \times M$ matrix, i.e., $J_x = J_y = J_t = 0$,  
satisfying $J^2 = I$. $\d$ and $\bd$ extend to matrices over $\cA$ and to the corresponding graded algebra.  
Setting $m=1$ and 
\be
    \Delta = \Gamma = - 2 I_n \, \pa \, , 
    \qquad   \kappa = \lambda = 0 \, , 
      \label{DS_Delta,Gamma,kappa,lambda}
\ee
the conditions (\ref{Delta,lambda,Gamma,kappa_eqs}) are satisfied.
(\ref{G.o=o.D}) becomes $\omega_x = 0$, and the second equation in (\ref{c,gamma_in_terms_of_omega}) yields
\bez
      \gamma = \omega_y \, \xi_1 - ( \imag \, \omega_t - 2 \, \omega_y \, \pa ) \, \xi_2   \, .
\eez
The linear equations (\ref{theta_eq}) and (\ref{eta_eq}) read
\be
    \theta_y = J \theta_x + \frac{1}{2} \, [J,\phi_0] \, \theta \, , \qquad
    \imag \, \theta_t = - 2 \, J \theta_{xx} - [J,\phi_0] \, \theta_x - (\phi_{0,y} + J \phi_{0,x}) \, \theta \, , \nonumber \\
    \eta_y = \eta_x \, J - \frac{1}{2} \, \eta \, [J,\phi_0] \, , \qquad
    \imag \, \eta_t = 2 \, \eta_{xx} \, J - \eta_x \, [J,\phi_0] + \eta \, (\phi_{0,y} + \phi_{0,x} \, J) \, . 
       \label{DS_linsys}
\ee
 From (\ref{scs_Omega_eqs}) we obtain
\be
  \Omega_x &=& - \frac{1}{2} \eta \, \theta \, , \qquad
  \Omega_y = - \frac{1}{2} \eta \, J \, \theta + \omega_y \, ,  \nonumber \\
  \imag \, \Omega_t &=& - \eta_x \, J \, \theta + \eta \, J \, \theta_x + \frac{1}{2} \eta \, [J , \phi_0 ] \, \theta
      + \imag \, \omega_t \, .    \label{DS_Omega_eqs}
\ee
Given solutions $\theta$ and $\eta$ of the linear equations, and a solution $\Omega$ of the latter 
consistent system, then 
\bez
    \phi = \phi_0 - \theta \, \Omega^{-1} \, \eta 
           \, , \qquad
    q = \theta \, \Omega^{-1}  \, , \qquad 
    r = \Omega^{-1} \, \eta 
\eez
solve  
\be
   \lefteqn{ \frac{\imag}{2} [J,\phi_t] + \phi_{yy} - J \phi_{xx} J + \frac{1}{2} [\phi_y+J\phi_x , [J,\phi]] 
    + \frac{1}{2} [J,\phi] [J,\phi_x] } \hspace{.5cm}  \nonumber \\
  &=& (q \, \omega_y \, r)_y + J \, (q \, \omega_y \, r)_x + \frac{\imag}{2} [J , q \, \omega_t \, r ]
      - [ J , q \, \omega_y \, r_x ] \, ,  \label{DS_scs}
\ee
which is obtained from (\ref{scs_phi_eq}), and 
\be
    && q_y = J q_x + \frac{1}{2} [J,\phi] \, q - q \, \omega_y \, \Omega^{-1}  \, , \nonumber \\
    && \imag \, q_t = - 2 J q_{xx}
       - [J,\phi] \, q_x - ( \phi_y + J \phi_x ) \, q    
       + q \, \omega_y \, r \, q - \imag \, q \, \omega_t \, \Omega^{-1}  \, ,   \nonumber \\
    && r_y = r_x \, J - \frac{1}{2} r \, [J,\phi] - \Omega^{-1} \, \omega_y \, r   \, ,  \nonumber \\
    && \imag \, r_t = 2 r_{xx} \, J 
         - r_x \, [J,\phi] + r \, ( \phi_y + \phi_x \, J  ) 
        - r \, q \, \omega_y \, r - \imag \, \Omega^{-1} \, \omega_t \, r \, ,  \label{DS_scs_lin} 
\ee
which results from (\ref{scs_q,r_eqs}). 
The equations (\ref{DS_scs}) and (\ref{DS_scs_lin}) include the following two systems with self-consistent sources. 
\begin{enumerate}
\item $\omega_y=0$. Then (\ref{DS_scs}) and two of the equations (\ref{DS_scs_lin}) read
\be
   && \frac{\imag}{2} [J,\phi_t] + \phi_{yy} - J \phi_{xx} J + \frac{1}{2} [\phi_y+J\phi_x , [J,\phi]] 
    + \frac{1}{2} [J,\phi] [J,\phi_x] 
     = \frac{\imag}{2} [J , q \, \omega_t \, r ]  \, ,  \nonumber   \\
   && q_y = J q_x + \frac{1}{2} [J,\phi] \, q \, , \qquad
      r_y = r_x \, J - \frac{1}{2} r \, [J,\phi] \, .     \label{DS_scs_1} 
\ee
\item $\omega_t=0$. In this case we obtain
\be
 \lefteqn{ \frac{\imag}{2} [J,\phi_t] + \phi_{yy} - J \phi_{xx} J + \frac{1}{2} [\phi_y+J\phi_x , [J,\phi]] 
    + \frac{1}{2} [J,\phi] [J,\phi_x] } \hspace{.5cm}  \nonumber \\
    &=& (q \, \omega_y \, r)_y + J \, (q \, \omega_y \, r)_x - [ J , q \, \omega_y \, r_x ] \, , \nonumber \\
  \imag \, q_t &=& - 2 J q_{xx} - [J,\phi] \, q_x - ( \phi_y + J \phi_x ) \, q + q \, \omega_y \, r \, q \, , \nonumber \\
  \imag \, r_t &=& 2 r_{xx} \, J - r_x \, [J,\phi] + r \, ( \phi_y + \phi_x \, J  ) - r \, q \, \omega_y \, r
       \, . \label{DS_scs_2} 
\ee
\end{enumerate}

\paragraph{Hermitian conjugation reductions.} 
(\ref{DS_scs}) and (\ref{DS_scs_lin}), and its special cases (\ref{DS_scs_1}) and (\ref{DS_scs_2}), are 
compatible with the reduction conditions (\ref{NLS_hc_reduction}). Using the same block-decomposition as in (\ref{NLS_block-decomp}),  
we have again the conditions (\ref{NLS_decomp_hc_reduction}), considered in Section~\ref{sec:NLS}. 
The last two equations in (\ref{DS_scs_lin}) are then redundant, and we obtain the following reduced systems with 
self-consistent sources. Without restriction of generality, we can set $\epsilon_1=1$. 
\begin{enumerate}
\item $\omega_y=0$. From (\ref{DS_scs_1}) we obtain the following matrix Davey-Stewartson (DS) 
equation with self-consistent sources:
\be
   && \imag \, \mathfrak{q}_t + \mathfrak{q}_{xx} + \mathfrak{q}_{yy}
         + (u_x + u_y) \, \mathfrak{q} - \mathfrak{q} \, (v_x - v_y)
      = - \imag \, \varepsilon \, q_1 \, \omega_t \, q_2^\dagger  \, ,       \nonumber    \\
   && u_{xx} - u_{yy} = \varepsilon \, (\mathfrak{q} \, \mathfrak{q}^\dagger)_x + \varepsilon \, (\mathfrak{q} \, \mathfrak{q}^\dagger)_y 
                \, , \qquad
      v_{xx} - v_{yy} = \varepsilon \, (\mathfrak{q}^\dagger \, \mathfrak{q})_y - \varepsilon \, (\mathfrak{q}^\dagger \, \mathfrak{q})_x 
                \, , \nonumber \\
   && q_{1,y} - q_{1,x} = \mathfrak{q} \, q_2  \, , \quad
      q_{2,y} + q_{2,x} = \varepsilon \,\mathfrak{q}^\dagger \, q_1     \, . \label{DS_scs_1_decomp}
\ee
\item $\omega_t=0$. From (\ref{DS_scs_2}) we obtain another matrix Davey-Stewartson (DS) 
equation with self-consistent sources:
\be
   && \imag \, \mathfrak{q}_t + \mathfrak{q}_{xx} + \mathfrak{q}_{yy}
         + (u_x + u_y) \, \mathfrak{q} - \mathfrak{q} \, (v_x - v_y)        \nonumber \\
   && \qquad   = - \varepsilon \, q_{1,x} \, \omega_y \, q_2^\dagger 
        + \varepsilon \, q_1 \, \omega_y \, q_{2,x}^\dagger 
        - \varepsilon \, (q_1 \, \omega_y \, q_2^\dagger)_y \, , \qquad \nonumber \\
   && u_{xx} - u_{yy} - \varepsilon \, (\mathfrak{q} \, \mathfrak{q}^\dagger)_x - \varepsilon \, (\mathfrak{q} \, \mathfrak{q}^\dagger)_y
      = - ( q_1 \, \omega_y \, q_1^\dagger)_x - ( q_1 \, \omega_y \, q_1^\dagger)_y    \, , \nonumber \\
   && v_{xx} - v_{yy} + \varepsilon \, (\mathfrak{q}^\dagger \, \mathfrak{q})_x - \varepsilon \,  (\mathfrak{q}^\dagger \, \mathfrak{q})_y
      = \varepsilon \, ( q_2 \, \omega_y \, q_2^\dagger)_x - \varepsilon \, ( q_2 \, \omega_y \, q_2^\dagger)_y 
              \, , \nonumber \\
   && \imag \, q_{1,t} + 2 \, q_{1,xx} + 2 \, \mathfrak{q} \, q_{2,x} + (\mathfrak{q}_x + \mathfrak{q}_y) \, q_2
       + (u_x + u_y) \, q_1 = q_1 \, \omega_y \, (q_1^\dagger \, q_1 - \varepsilon \, q_2^\dagger \, q_2 ) \, , \nonumber \\
   && \imag \, q_{2,t} - 2 \, q_{2,xx} + 2 \varepsilon \, \mathfrak{q}^\dagger \, q_{1,x} 
      + \varepsilon \, ( \mathfrak{q}_x^\dagger - \mathfrak{q}_y^\dagger ) \, q_1
      + (v_x - v_y) \, q_2 = q_2 \, \omega_y \, ( q_1^\dagger \, q_1 - \varepsilon \, q_2^\dagger \, q_2 ) \, . 
      \qquad \quad           \label{DS_scs_2_decomp}
\ee
\end{enumerate}
In order that the solution generating procedure respects the reduction conditions, we also have to require 
\be
      \eta = \theta^\dagger \Lambda \, ,   \label{DS_theta_eta_red}
\ee
with $\Lambda$ defined in (\ref{NLS_hc_reduction}). 

\begin{remark}
In the scalar case, i.e., $M_1=M_2=1$, the inhomogeneous linear equations for $u$ and $v$ in (\ref{DS_scs_1_decomp}) 
integrate to $u_y - u_x = - \varepsilon \, |\mathfrak{q}|^2 = v_y + v_x$, 
dropping ``constants'' of integration. This is solved by
$u = w_y + w_x$ and $v = w_y - w_x$, with a function $w$, and we obtain $w_{xx} - w_{yy} = \varepsilon \, |\mathfrak{q}|^2$. 
The first of equations (\ref{DS_scs_1_decomp}) then takes the form
$\imag \, \mathfrak{q}_t + \mathfrak{q}_{xx} + \mathfrak{q}_{yy}
         + 2 \, (w_{xx} + w_{yy}) \, \mathfrak{q} = - \imag \, \varepsilon \, q_1 \, \omega_t \, q_2^\dagger$. 
Passing over to ``light cone variables'',
we recover the DS equation with self-consistent sources treated in \cite{HWT08} (also see \cite{Shen+Jian09}).
\end{remark}

\begin{remark}
\label{DS_Scalar_remark} 
We consider again the scalar case $M_1=M_2=1$. Introducing the new variable 
$s = u + v + \int^y ( u_x - v_x - 2 \varepsilon \, |\mathfrak{q}|^2\,) \, dy$, 
 from (\ref{DS_scs_1_decomp}) we obtain
\be
   && \imag \, \mathfrak{q}_t +  \mathfrak{q}_{xx} + \mathfrak{q}_{yy}  + s_y \, \mathfrak{q}
     + 2 \varepsilon \, |\mathfrak{q}|^2 \mathfrak{q} = - \imag \, q_1 \, \omega_t \, q_2^\dagger  \, ,  \nonumber  \\
   && s_{xx} - s_{yy} = 4 \varepsilon \, (|\mathfrak{q}|^2)_y + h(x) \, ,    \nonumber \\
   && q_{1,y} - q_{1,x} = \mathfrak{q} \, q_2  \, , \quad
      q_{2,y} + q_{2,x} = \varepsilon \,\mathfrak{q}^\dagger \, q_1  \, .      \label{scalarDS_Scs} 
\ee
Here $h(x)$ is an arbitrary function of $x$, which can be eliminated by a redefinition of $s$. 
In this way we recover the system of Example~3 in \cite{HLYZ11}.
\end{remark}

\begin{remark}
Assuming that all objects do not depend on the variable $x$, (\ref{DS_scs_1_decomp}) reduces to the NLS equation 
in Section~\ref{sec:NLS} up to changes in notation ($J \mapsto -J$, and $y$ has finally to be renamed to $x$). 
We note that the second of equations (\ref{DS_Omega_eqs}) becomes a constraint: $\eta \, J \, \theta = 0$. As formulated 
above, our solution generating method does not work for this NLS reduction. This can be corrected by   
extending the first equations in (\ref{DS_Delta,Gamma,kappa,lambda}) to $\Delta = - 2 I_n \, \pa + P$ and 
$\Gamma = - 2 I_n \, \pa + Q$, with constant $n \times n$ matrices $P,Q$, and generalizing the subsequent equations 
accordingly. In particular, we need non-vanishing $\kappa$ and $\lambda$, cf. Section~\ref{sec:NLS}. 
The matrices $P$ and $Q$ are redundant on the level of DS. 

Dropping in (\ref{DS_bidiff}) the partial derivatives with respect to $x$, we recover the bidifferential calculus 
for the NLS system in Section~\ref{sec:NLS}, up to the stated changes in notation. Dropping instead the partial derivatives 
with respect to $y$, we obtain a bidifferential calculus for the NLS system different from that used in Section~\ref{sec:NLS}.
\end{remark}

\paragraph{Exact solutions in case of vanishing seed.}
Let $\phi_0 = 0$. 
The linear equations (\ref{DS_linsys}) are solved by
\bez
   \theta = a_+ \, e^{\vartheta_1(P_1)} + a_- \, e^{\vartheta_2(P_2)} \, , \quad 
   \eta = e^{-\vartheta_1(Q_1)} \, b_+ + e^{-\vartheta_2(Q_2)} \, b_- \, , 
\eez
with constant matrices $a_\pm, b_\pm, P_i,Q_i$, $i=1,2$, and 
\bez
  J \, a_{\pm} = \pm a_{\pm} \, , \quad 
  b_\pm \, J = \pm b_\pm \, , \quad 
  \vartheta_1(P) = P \, (x + y) + 2 \imag \, P^2 \, t  \, , \quad
  \vartheta_2(P) = P \, (y - x) - 2 \imag \, P^2 \, t   \, .
\eez   
(\ref{DS_theta_eta_red}) requires
\bez 
      b_+ = a_+^\dagger \, , \quad
      b_- = - \varepsilon \, a_-^\dagger \, , \quad
      Q_i = - P_i^\dagger \qquad i=1,2  \, .
\eez      
A corresponding solution of (\ref{NLS_Omega_eqs}) is
\bez
    \Omega = e^{-\vartheta_1(-P_1^\dagger)} \, A \, e^{\vartheta_1(P_1)} + e^{-\vartheta_2(-P_2^\dagger)} \, B \, e^{\vartheta_2(P_2)}
             + \omega \, ,
\eez
where the constant matrices $A, B$ have to satisfy the Sylvester equations
\bez
   P_1^\dagger \, A + A \, P_1 = - \frac{1}{2} \, a_+^\dagger a_+ \, , \qquad 
   P_2^\dagger \, B + B \, P_2 = - \frac{1}{2} \, \varepsilon \, a_-^\dagger a_- \, .
\eez
If $A^\dagger = A$ and $B^\dagger = B$, then $\Omega^\dagger = \Omega$ and $\phi = - \theta \, \Omega^{-1} \theta^\dagger \, \Lambda$ 
satisfies the reduction conditions. Together with $q = \theta \, \Omega^{-1}$, after decomposition we then obtain solutions of 
(\ref{DS_scs_1_decomp}) and (\ref{DS_scs_2_decomp}), if $\omega_y=0$, respectively $\omega_t=0$.

\begin{example}
Let $M_1=M_2=1$ and $n=1$. We choose $P_i = \lambda_i \, I$ and 
\bez
    a_+ = \left( \begin{array}{cc} 1 & 0 \\ 0 & 0 \end{array} \right) \, , \quad
    a_- = \left( \begin{array}{cc} 0 & 0 \\ 0 & 1 \end{array} \right) \, , \quad
    \omega = \left( \begin{array}{cc} \omega_{11} & \omega_{12} \\ \omega_{12}^\ast & \omega_{22} \end{array} \right) \, ,
\eez
where $^\ast$ denotes complex conjugation. The Sylvester equations are easily solved and we obtain
\bez
    \Omega = \left( \begin{array}{cc} \omega_{11} - \frac{1}{4 \, \mathrm{Re}(\lambda_1)} \, e^{\alpha_1}  & \omega_{12} \\ 
             \omega_{12}^\ast & \omega_{22} - \frac{1}{4 \, \mathrm{Re}(\lambda_2)} \, e^{\alpha_2}  \end{array} \right) \, ,
\eez
where 
\bez
   \alpha_1 = 2 \, \mathrm{Re}(\lambda_1) \, (x+y) - 8 \, \mathrm{Re}(\lambda_1) \, \mathrm{Im}(\lambda_1) \, t \, , \quad 
   \alpha_2 = 2 \, \mathrm{Re}(\lambda_2) \, (y-x) + 8 \, \mathrm{Re}(\lambda_2) \, \mathrm{Im}(\lambda_2) \, t \, . 
\eez   
Hence
\bez
    \det(\Omega) = \omega_{11} \, \omega_{22} - |\omega_{12}|^2 - \frac{\omega_{22}}{4 \, \mathrm{Re}(\lambda_1)} \, e^{\alpha_1} 
                  - \frac{\omega_{11}}{4 \, \mathrm{Re}(\lambda_2)} \, e^{\alpha_2}
                  + \frac{1}{4 \, \mathrm{Re}(\lambda_1) \, \mathrm{Re}(\lambda_2)} \, e^{\alpha_1 + \alpha_2} \, .
\eez
Evaluation of $\phi = - \theta \, \Omega^{-1} \theta^\dagger \, \Lambda$ and its decomposition now leads to 
\bez
  &&  \mathfrak{q} = - \frac{\varepsilon \, \omega_{12}}{\det(\Omega)} \, e^{(\lambda_1 - \lambda_2^\ast ) \, x 
                        + (\lambda_1+\lambda_2^\ast) \, y
                        + 2 \imag \, ( \lambda_1^2+ (\lambda_2^\ast)^2 ) \, t} \, ,  \\
 && u= - \frac{1}{\det(\Omega)} \, e^{\alpha_1} \, 
      \Big( \omega_{22} - \frac{1}{4 \, \mathrm{Re}(\lambda_2)} \, e^{\alpha_2} \Big) \, , \quad
    v = - \frac{\varepsilon}{\det(\Omega)} \, e^{\alpha_2} \, \Big( \omega_{11} - \frac{1}{4 \, \mathrm{Re}(\lambda_1)} \, e^{\alpha_1} \Big) \, .
\eez
 From $q = \theta \, \Omega^{-1}$ we obtain
\bez
   && q_1 = \frac{1}{\det(\Omega)} \, e^{\vartheta_1(\lambda_1)} \,
           ( \omega_{22} - \frac{1}{4 \, \mathrm{Re}(\lambda_2)} \, e^{\alpha_2}  \, , \, - \omega_{12} )  \, , \\
   && q_2 = \frac{1}{\det(\Omega)} \, e^{\vartheta_1(\lambda_2)} \, 
           ( - \omega_{12}^\ast \, , \, \omega_{11} - \frac{1}{4 \, \mathrm{Re}(\lambda_1)} \, e^{\alpha_1} ) \, .
\eez
If $\det(\Omega)$ nowhere vanishes, this solution is regular. If $\omega$ is constant, it describes a single dromion solution 
of the DS equation \cite{BLMP88,Foka+Sant89,Hiro+Hiet90,Gils+Nimm91}, or its degeneration to a solitoff \cite{Gils92} or a soliton. 
A solution of a DS equation with self-consistent sources, with non-constant $\omega$, can change its type. In particular, we find
the following.
\begin{enumerate}
\item In a region where $\det(\omega)<0$, $\omega_{11}\, \mathrm{Re}(\lambda_2) >0$, $\omega_{22}\,\mathrm{Re}(\lambda_1)>0$,
$\mathrm{Re}(\lambda_1) \, \mathrm{Re}(\lambda_2)<0$, the above solution describes a single dromion \cite{Hiro+Hiet90}.
\item If $\det(\omega)=0$, $\omega_{11}\, \mathrm{Re}(\lambda_2)>0$, $\omega_{22}\, \mathrm{Re}(\lambda_1)>0$,
$\mathrm{Re}(\lambda_1) \, \mathrm{Re}(\lambda_2)<0$, we have a single solitoff solution \cite{Gils92}.
\item If $\omega_{11}=\omega_{22}=0$, $\mathrm{Re}(\lambda_1) \, \mathrm{Re}(\lambda_2)<0$, the solution represents 
a single soliton.
\end{enumerate}

A nonlinear superposition of $n$ of such elementary solutions is obtained by taking $a_\pm$ to be a row of $n$ $2 \times 2$ matrices of 
the above form, and $P_i = \mbox{block-diagonal}(\lambda_{i1} \, I_2, \ldots, \lambda_{in} \, I_2)$. 
\end{example}

\section{Matrix two-dimensional Toda lattice equation with self-consistent sources}
\label{sec:Toda}
\setcounter{equation}{0}
Let $\cA_0$ be the space of complex functions on $\mathbb{R}^2 \times \mathbb{Z}$, smooth in the 
first two variables. 
We extend it to $\cA = \cA_0[\bbS,\bbS^{-1}]$, where $\bbS$ is the shift operator in the discrete variable $k \in \bbZ$.
A bidifferential calculus is determined by setting 
\bez
    \d f = [\bbS,f] \, \xi_1 + [\pa_y,f] \, \xi_2 \, , \qquad
    \bd f = [\pa_x,f] \, \xi_1 - [\bbS^{-1},f] \, \xi_2 
\eez
on $\cA$ \cite{DMH08bidiff}. For $f \in \cA$, we write $f^{\pm}(x,y,k) := f(x,y,k \pm 1)$.  
We set 
\bez
      \phi = \varphi \, \bbS^{-1} \, , \quad 
         q = \tilde{q} \, \bbS^{-1} \, , \quad
         r = \bbS^{-1} \, \tilde{r} \, , \quad
      \Delta = \Gamma = \bbS^{-1} \, ,  \quad 
      \kappa = \lambda = 0 \, , \quad
      \Omega = \tilde{\Omega} \, \bbS \, , \quad 
      \omega = \tilde{\omega} \, \bbS \, . 
\eez
Then (\ref{Delta,lambda,Gamma,kappa_eqs}) is satisfied, (\ref{G.o=o.D}) becomes $\tilde{\omega}^+ = \tilde{\omega}$, 
and the second equation in (\ref{c,gamma_in_terms_of_omega}) leads to
\bez
      \gamma = \tilde{\omega}_x \, \bbS \, \xi_1 - \tilde{\omega}_y \, \xi_2   \, .
\eez
Now (\ref{scs_phi_eq}) becomes\footnote{For vanishing right hand side, this equation already appeared in \cite{DMH08bidiff}.}
\be
  \varphi_{xy} - (\varphi^+ - \varphi) \, (\varphi_y + I) + (\varphi_y + I) \, (\varphi - \varphi^-) 
  = ( \tilde{q} \, \tilde{\omega}_x \, \tilde{r}^- )_y 
    + [ \tilde{q}^+ \, \tilde{\omega}_y \, \tilde{r}^- - ( \tilde{q}^+ \, \tilde{\omega}_y \, \tilde{r}^- )^- ]
        \, .  \label{dToda_scs_eq}
\ee

\begin{remark}
In the scalar case, in terms of $V := \varphi_y$, (\ref{dToda_scs_eq}) without sources becomes 
$(\ln(1+V))_x = \varphi^+ - \varphi + \varphi^-$. Differentiating with respect to $y$, this becomes
the \emph{two-dimensional Toda lattice equation} $(\ln(1+V))_{xy} = V^+ - 2V + V^-$ \cite{Mikh79,HIK88,Hiro04}. 
A self-consistent source extension, and corresponding exact solutions, has been obtained in \cite{WHG07} 
(also see \cite{Hu+Wang06}), using the framework of Hirota's bilinear difference operators \cite{Hiro04}. 
(\ref{dToda_scs_eq}) leads to a matrix version. 
\end{remark}

Let $\varphi_0$ be a solution of (\ref{dToda_scs_eq}) with vanishing sources. 
The linear equations (\ref{theta_eq}) and (\ref{eta_eq}) read
\be
  && \theta_x = \theta^+ - \theta + (\varphi_0^+ - \varphi_0) \, \theta \, , \qquad
     \theta_y = \theta - \theta^- - \varphi_{0,y} \, \theta^- \, , \nonumber \\
  && \eta_x = \eta - \eta^- - \eta \, (\varphi_0^+ - \varphi_0) \, , \qquad
     \eta_y = \eta^+ - \eta + \eta^+ \, \varphi_{0,y}^+ \, ,  \label{Toda_lin_eqs}
\ee
and from (\ref{scs_Omega_eqs}) we obtain  
\be
     \tilde{\Omega} - \tilde{\Omega}^- = - \eta \, \theta  \,  ,   \qquad
     \tilde{\Omega}_x = - \eta \, \theta^+ + \tilde{\omega}_x \, , \qquad
     \tilde{\Omega}_y = - \eta^+ \, (\varphi_{0,y}^+ + I) \, \theta + \tilde{\omega}_y \, .    \label{Toda_Omega_eqs}
\ee
If $\theta, \eta$ and $\tilde{\Omega}$ solve the preceding equations, then 
\be
      \varphi = \varphi_0 - \theta \, (\tilde{\Omega}^-)^{-1} \, \eta^- 
      \, ,  \qquad
      \tilde{q} = \theta \, (\tilde{\Omega}^-)^{-1} \, , \qquad
      \tilde{r} = \tilde{\Omega}^{-1} \, \eta     \label{Toda_phi,q,r}
\ee
satisfy (\ref{dToda_scs_eq}) and the system
\be
 \tilde{q}_x &=& \tilde{q}^+ -\tilde{q}  + (\varphi^+ - \varphi) \, \tilde{q} 
                          -\tilde{q} \, \tilde{\omega}_x \,  (\tilde{\Omega}^-)^{-1} \, ,  \nonumber\\
 \tilde{q}_y  &=& \tilde{q} - \tilde{q}^- - \varphi_y \, \tilde{q}^-
                          - \tilde{q} \, \tilde{\omega}_y \,  (\tilde{\Omega}^{--})^{-1}  \, ,   \nonumber  \\
 \tilde{r}_x &=& \tilde{r} - \tilde{r}^{-}  - \tilde{r} \, (\varphi^+ - \varphi) 
                          - \tilde{\Omega}^{-1} \, \tilde{\omega}_x \, \tilde{r} \, , \nonumber \\
  \tilde{r}_y   &=& \tilde{r}^+ - \tilde{r} + \tilde{r}^+ \, \varphi_y^+  
                             - (\tilde{\Omega}^+)^{-1} \, \tilde{\omega}_y \, \tilde{r}  \, ,
              \label{dToda_scs_lin_sys}
\ee
which results from (\ref{scs_q,r_eqs}).

Using the extended Miura transformation, 
\bez
   \varphi^+ - \varphi = g_x \, g^{-1} - \tilde{q} \, \tilde{\omega}_x \, \tilde{r} \, , \qquad
   \varphi_y + I = ( g^+ \, g^{-1} + \tilde{q}^+ \, \tilde{\omega}_y \, \tilde{r} )^-   \,  ,
\eez
the system (\ref{dToda_scs_lin_sys}) becomes
\be
  \tilde{q}_x &=& \tilde{q}^+ - \tilde{q}  + ( g_x \, g^{-1} - \tilde{q} \, \tilde{\omega}_x \, \tilde{r} ) \, \tilde{q} 
                            - \tilde{q} \, \tilde{\omega}_x \, (\tilde{\Omega}^-)^{-1}     \, ,    \nonumber \\
 \tilde{q}_y  &=& \tilde{q} - ( g^+ \, g^{-1} + \tilde{q}^+ \, \tilde{\omega}_y \,  \tilde{r} )^- \, \tilde{q}^- 
                              - \tilde{q}  \, \tilde{\omega}_y \, (\tilde{\Omega}^{--})^{-1}    \, ,   \nonumber  \\
 \tilde{r}_x &=& \tilde{r} - \tilde{r}^{-}  - \tilde{r} \, ( g_x \, g^{-1} - \tilde{q} \, \tilde{\omega}_x \,\tilde{r} )
                          - \tilde{\Omega}^{-1} \, \tilde{\omega}_x \, \tilde{r}       \, , \nonumber \\
     \tilde{r}_y &=& - \tilde{r} + \tilde{r}^+ \, ( g^+ \, g^{-1} + \tilde{q}^+ \, \tilde{\omega}_y \, \tilde{r} )
                          - (\tilde{\Omega}^+)^{-1} \, \tilde{\omega}_y\, \tilde{r}  \, ,  \label{Toda_scs_lin_sys}
\ee
and (\ref{scs_g_eq}) reads\footnote{In the scalar case, writing $g = e^u$, the left hand side of 
(\ref{Toda_scs_eq}) becomes $u_{xy} - e^{u^+ - u} + e^{u-u^-}$.  }
\be
     (g_x \, g^{-1})_y - [ g^+ \, g^{-1} - (g^+ \, g^{-1})^- ]
  = + (\tilde{q} \, \tilde{\omega}_x \, \tilde{r})_y 
    + [ \tilde{q}^+ \, \tilde{\omega}_y \, \tilde{r} - (\tilde{q}^+ \, \tilde{\omega}_y \, \tilde{r})^- ]
              \, .        \label{Toda_scs_eq}
\ee
This system is solved by  
\be
     g = I - \theta \, ({\tilde{\Omega}}^{-})^{-1} \, \eta \, ,   \label{Toda_g}
\ee
together with $\tilde{q}$ and $\tilde{r}$ given by (\ref{Toda_phi,q,r}). 

Equations (\ref{dToda_scs_eq}) and (\ref{dToda_scs_lin_sys}), 
respectively (\ref{Toda_scs_lin_sys}) and (\ref{Toda_scs_eq}),
lead to versions of a two-dimensional matrix Toda lattice equation with 
self-consistent sources in the following cases.  

\begin{enumerate}
\item  $\tilde{\omega}_x=0$. (\ref{dToda_scs_eq}) reduces to
\be
 \varphi_{xy} - (\varphi^+ - \varphi) \, (\varphi_y + I)
                   + (\varphi_y + I) \, (\varphi - \varphi^-)
  = \tilde{q}^+ \, \tilde{\omega}_y \,\tilde{r}^- - ( \tilde{q}^+ \, \tilde{\omega}_y \, \tilde{r}^- )^-    \, . \label{dTodaSCS1}
\ee
Those of the equations (\ref{dToda_scs_lin_sys}) that do not involve $\tilde{\Omega}$ are
\be
 && \tilde{q}_x = \tilde{q}^+ - \tilde{q} + (\varphi^+ - \varphi) \, \tilde{q} \, , \qquad 
  \tilde{r}_x = \tilde{r} - \tilde{r}^{-} - \tilde{r} \, (\varphi^+ - \varphi)  \, . \label{dTodaSCSlin1}
\ee
(\ref{dTodaSCS1}) and (\ref{dTodaSCSlin1}) constitute the first type of the two-dimensional 
matrix Toda lattice equation with self-consistent sources. 

Correspondingly, (\ref{Toda_scs_eq}) and (\ref{Toda_scs_lin_sys}) reduce to
\be
&& (g_x \, g^{-1})_y - [ g^+ \, g^{-1} - (g^+ \,  g^{-1})^- ]
 = \tilde{q}^+ \, \tilde{\omega}_y \, \tilde{r} - ( \tilde{q}^+ \, \tilde{\omega}_y \, \tilde{r} )^-   \, , \nonumber \\
&& \tilde{q}_x = \tilde{q}^+ - \tilde{q} + g_x \, g^{-1} \, \tilde{q} \, , \qquad 
  \tilde{r}_x = \tilde{r} - \tilde{r}^{-}  - \tilde{r} \, g_x \, g^{-1}  \, .   \label{TodaSCS1}
\ee
These equations constitute the Miura-dual of the first type of the two-dimensional matrix Toda lattice equation
with self-consistent sources. In the scalar case, in terms of $u = \ln g$ this takes the form
\bez
    && u_{xy} - e^{u^+-u} + e^{u-u^-} = \tilde{q}^+ \, \tilde{\omega}_y \, \tilde{r} 
            - ( \tilde{q}^+ \, \tilde{\omega}_y \, \tilde{r} )^- \, ,  \\
    && \tilde{q}_x = \tilde{q}^+ - \tilde{q} + u_x \, \tilde{q} \, , \qquad 
       \tilde{r}_x = \tilde{r} - \tilde{r}^{-}  - u_x \, \tilde{r}  \, .  
\eez
\item  $\tilde{\omega}_y=0$. (\ref{dToda_scs_eq}) becomes
\be
 \varphi_{xy} - (\varphi^+ - \varphi) \, (\varphi_y + I) + (\varphi_y + I) \, (\varphi - \varphi^-)  
 =  (\tilde{q} \, \tilde{\omega}_x \, \tilde{r}^- )_y \, .
  \label{dTodaSCS2}
\ee
Those of equations (\ref{dToda_scs_lin_sys}) that do not depend on $\tilde{\Omega}$ are
\be
  \tilde{q}_y  = \tilde{q} - \tilde{q}^- - \varphi_y \, \tilde{q}^- \, ,   \qquad     
  \tilde{r}_y = \tilde{r}^+ - \tilde{r} +\tilde{r}^+ \, \varphi_y^+ \, . \label{dTodaSCSlin2}
\ee
(\ref{dTodaSCS2}) and (\ref{dTodaSCSlin2}) constitute the second type of the two-dimensional 
matrix Toda lattice equation with self-consistent sources. 

 From (\ref{Toda_scs_eq}) and (\ref{Toda_scs_lin_sys}) we obtain the corresponding Miura-dual,  
\be
  && (g_x \, g^{-1})_y - [ g^+ \, g^{-1} - (g^+ \,  g^{-1})^- ] 
   = (\tilde{q} \, \tilde{\omega}_x \,\tilde{r})_y \, , \nonumber \\
  && \tilde{q}_y = \tilde{q} - g \, (g^{-1} \, \tilde{q})^- \, ,  \qquad
    \tilde{r}_y = -\tilde{r} +\tilde{r}^+ \, g^+ \, g^{-1}   \, .  \label{TodaSCS2}
\ee
In the scalar case ($m=1$), in terms of $u = \ln g$,
$a = e^{-y} \, g^{-1} \, \tilde{q} \, \tilde{\omega}_x$ and
$b = e^{y} \, g \, \tilde{r}$, (\ref{TodaSCS2}) can be expressed as follows, 
\bez
       u_{xy} - e^{u^+-u} + e^{u-u^-} = (a \, b)_y \, ,  \quad
       a_y + u_y \, a + a^- = 0 \, ,  \quad
       b_y - b \, u_y - b^+ = 0 \, . 
\eez
Such a system appeared in \cite{LZL08Toda}.
\end{enumerate}

The above matrix versions of the two-dimensional Toda lattice equations with self-consistent sources 
are new according to our knowledge.

\paragraph{Explicit solutions for trivial seed.}
Let $\varphi_0 =0$ and $g_0 =I$. Then (\ref{Toda_lin_eqs}) becomes
\bez
  \theta_x = \theta^+ - \theta \, , \qquad
  \theta_y = \theta - \theta^- \, , \qquad
  \eta_x = \eta - \eta^- \, , \qquad
  \eta_y = \eta^+ - \eta \, .
\eez
Special solutions are given by
\bez
  && \theta = a \, e^{\vartheta(P)} \, P^k \, A  \, , \quad \eta = B \, e^{-\vartheta(Q)} \, Q^{-k}\, b, 
      \, \quad
     \vartheta(P) := (P -I) \, x + (I - P^{-1}) \, y \, ,  
\eez
with arbitrary constant matrices $a,b,A,B,P,Q$ of appropriate size.
(\ref{Toda_Omega_eqs}) leads to
\bez
    {\tilde{\Omega}} = \tilde{\omega} - B \, e^{-\vartheta(Q)} \, Q^{-k}\, X \, e^{\vartheta(P)} \, P^{k+1} \, A \, ,
\eez
where the constant matrix $X$ has to satisfy the Sylvester equation
\bez
    X \, P - Q \, X = b \, a \, .
\eez
Then (\ref{Toda_phi,q,r}) and (\ref{Toda_g}) provide us with explicit solutions of the above matrix 
two-dimensional Toda lattice equations with self-consistent sources.

\section{A generalized discrete KP equation with self-consistent sources}
\label{sec:dKP}
\setcounter{equation}{0}
Let $\cA_0$ be the space of complex functions of discrete variables $k_0,k_1,k_2 \in \bbZ$, and 
$\bbS_0,\bbS_1,\bbS_2$ corresponding shift operators. 
We extend $\cA_0$ to $\cA = \cA_0[\bbS_0^{\pm 1},\bbS_1^{\pm 1},\bbS_2^{\pm 1}]$ and define $\d$ 
and $\bar{\d}$ on $\cA$ via 
\be
  \d f = \sum_{i=1}^2 c_i^{-1} \, [ \bbS_i^{-1} , f] \, \xi_i \, , \qquad
  \bar{\d} f = \sum_{i=1}^2 [ \bbS_i^{-1} \bbS_0 , f] \, \xi_i \, ,
        \label{HM_bidiff}
\ee 
where $c_i$ are constants. Then $\d$ and $\bar{\d}$ extend to 
$\boldsymbol{\Omega} = \cA \otimes \bsy{\Lambda}$ and 
to matrices over $\boldsymbol{\Omega}$. In the following we will use the notation
\bez
    f_{,0} := \bbS_0 \, f \, \bbS_0^{-1} \, , \quad
    f_{,-0} := \bbS_0^{-1} \, f \, \bbS_0 \, ,  \quad
    f_{,i} := \bbS_i \, f \, \bbS_i^{-1} \, , \quad
    f_{,-i} := \bbS_i^{-1} \, f \, \bbS_i \qquad i=1,2 \, .
\eez 
We set 
\bez
      \Delta = \Gamma = \bbS_0 \, , \quad  
      \kappa = \lambda = 0 \, , \quad
      \Omega = \tilde{\Omega} \, \bbS_0^{-1} \, , \quad 
      \omega = \tilde{\omega} \, \bbS_0^{-1} \, , \quad 
      \phi = \varphi \, \bbS_0 \, , \quad 
      q = \tilde{q} \, \bbS_0 \, , \quad 
      r = \bbS_0 \, \tilde{r}  \, .
\eez
Then (\ref{Delta,lambda,Gamma,kappa_eqs}) is satisfied, (\ref{G.o=o.D}) becomes 
$\tilde{\omega}_{,0} = \tilde{\omega}$, 
and the second equation in (\ref{c,gamma_in_terms_of_omega}) leads to
\bez
     \gamma = \sum_{i=1}^2 \bbS_i^{-1} \, \gamma_i \,  \xi_i \, , \qquad
     \gamma_i := (c_i^{-1}-1) (\tilde{\omega}_{,i} - \tilde{\omega}) \, .
\eez
The linear equations (\ref{theta_eq}) and (\ref{eta_eq}) read 
\bez
 \theta_{,i} &=& (c_i-1)^{-1} \, [ (c_i + \varphi_{0,i} - \varphi_0) \, \theta_{,0} - \theta ] \, , \\
 \eta_{,-i} &=& (c_i-1)^{-1} \, [ \eta_{,-0} \, (c_i + \varphi_0 - \varphi_{0,-i})_{,-0} - \eta ] \, , 
\eez
and from (\ref{scs_Omega_eqs}) we obtain 
\bez
    \tilde{\Omega}_{,0} - \tilde{\Omega} = \eta \, \theta \, , \qquad
   \tilde{\Omega} - \tilde{\Omega}_{,-i} 
   = (c_i-1)^{-1} \, \eta_{,-0} \, \left( c_i + \varphi_0 - \varphi_{0,-i} \right)_{,-0} \, \theta_{,-i}
     - \tilde{\omega} + \tilde{\omega}_{,-i} \, .
\eez
Given solutions $\theta, \eta$ and $\tilde{\Omega}$, according to Section~\ref{sec:scs_via_bidiff}, 
\be
     \varphi = \varphi_0 - \theta \, \tilde{\Omega}_{,0}^{-1} \, \eta_{,0} \, , \qquad
     \tilde{q} = \theta \, \tilde{\Omega}_{,0}^{-1} \, , \qquad
     \tilde{r} = \tilde{\Omega}^{-1} \, \eta    \label{dKP_solutions}
\ee
solve the equations 
\bez
   \lefteqn{ (c_j + \varphi_{,j} - \varphi)_{,i} \, (c_i + \varphi_{,i} - \varphi)_{,0}
  - (c_i + \varphi_{,i} - \varphi)_{,j} \, (c_j + \varphi_{,j} - \varphi)_{,0}  } \hspace{2cm} \\
  &=& (c_i-1) \, [ (\tilde{q}_{,i} \, (\tilde{\omega}_{,i} - \tilde{\omega}) \, \tilde{r}_{,0,0})_{,j}
     - \tilde{q}_{,i} \, (\tilde{\omega}_{,i} - \tilde{\omega}) \, \tilde{r}_{,0,0} ]  \\
  && - (c_j-1) \, [(\tilde{q}_{,j} \, (\tilde{\omega}_{,j} - \tilde{\omega}) \, \tilde{r}_{,0,0})_{,i}
     - \tilde{q}_{,j} \, (\tilde{\omega}_{,j} - \tilde{\omega}) \, \tilde{r}_{,0,0} ]   \qquad \quad i,j=1,2
\eez
and 
\bez
  \tilde{q}_{,i} &=& (c_i-1)^{-1} [ ( c_i + \varphi_{,i} - \varphi) \, \tilde{q}_{,0} - \tilde{q} ] 
     - \tilde{q}_{,i} \, (\tilde{\omega}_{,i} - \tilde{\omega}) \, \tilde{\Omega}^{-1}_{,0,0} \, , \\
  \tilde{r}_{,-i} &=& (c_i-1)^{-1} [ \tilde{r}_{,-0} \, ( c_i + \varphi - \varphi_{,-i} )_{,-0} - \tilde{r} ]
     + \tilde{\Omega}_{,0}^{-1} \, (\tilde{\omega} - \tilde{\omega}_{,-i}) \, \tilde{r}_{,-i} \, ,  \qquad i=1,2 \, .
\eez
They result, respectively, from (\ref{scs_phi_eq}) and (\ref{scs_q,r_eqs}). 
The extended Miura equation (\ref{ext_Miura}) takes the form
\bez
    \varphi_{,i} - \varphi + c_i = c_i \, g_{,i} \, g_{,0}^{-1} 
      - (c_i-1) \, \tilde{q}_{,i} \, (\tilde{\omega}_{,i} - \tilde{\omega}) \, \tilde{r}_{,0}  \qquad \quad i=1,2 \, .
\eez
Correspondingly, (\ref{scs_g_eq}) becomes
\bez
  \lefteqn{  c_i \, \Big( ( g_{,i} \, g_{,0}^{-1} )_{,j} - g_{,i} \, g_{,0}^{-1} \Big)  
         - c_j \, \Big( ( g_{,j} \, g_{,0}^{-1} )_{,i} - g_{,j} \, g_{,0}^{-1} \Big) } \hspace{0cm}  \\
  &=& (c_i-1) \, [ ( \tilde{q}_{,i} \, (\tilde{\omega}_{,i} - \tilde{\omega}) \, \tilde{r}_{,0} )_{,j}
     - \tilde{q}_{,i} \, (\tilde{\omega}_{,i} - \tilde{\omega}) \, \tilde{r}_{,0} ]
     - (c_j-1) \, [ ( \tilde{q}_{,j} \, (\tilde{\omega}_{,j} - \tilde{\omega}) \, \tilde{r}_{,0} )_{,i}
     - \tilde{q}_{,j} \, (\tilde{\omega}_{,j} - \tilde{\omega}) \, \tilde{r}_{,0} ]  \, ,
\eez
and the above equations for $\tilde{q}$ and $\tilde{r}$ transform to
\bez
  \tilde{q}_{,i} &=& (c_i-1)^{-1} \, [ c_i \, g_{,i} \, g_{,0}^{-1} \, \tilde{q}_{,0} - q ] 
      - [ \tilde{q}_{,i} \, (\tilde{\omega}_{,i} - \tilde{\omega}) \, \tilde{r}_{,0} ] \, \tilde{q}_{,0}
      - \tilde{q}_{,i} \, (\tilde{\omega}_{,i} - \tilde{\omega}) \, \tilde{\Omega}^{-1}_{,0,0}  \, , \\
  \tilde{r}_{,-i} &=& (c_i-1)^{-1} \, [ c_i \, \tilde{r}_{,-0} \, g_{,-0} \, g_{,-i}^{-1} - \tilde{r} ] 
     - \tilde{r}_{,-0} \, [ \tilde{q}_{,-0} \, (\tilde{\omega} - \tilde{\omega}_{,-i})_{,-0} \, \tilde{r}_{,-i} ]
     + \tilde{\Omega}_{,0}^{-1} \, (\tilde{\omega} - \tilde{\omega}_{,-i}) \, \tilde{r}_{,-i} \, .
\eez

Setting $\tilde{\omega}_{,1} = \tilde{\omega}$, and retaining only those equations for $\tilde{q}$ 
and $\tilde{r}$ that do not contain $\tilde{\Omega}$, we obtain the following equations with self-consistent sources, 
\bez
  \lefteqn{  (c_1 + \varphi_{,1} - \varphi)_{,2} \, (c_2 + \varphi_{,2} - \varphi)_{,0}
  - (c_2 + \varphi_{,2} - \varphi)_{,1} \, (c_1 + \varphi_{,1} - \varphi)_{,0}   } \hspace{1cm} \nonumber \\
  &=& (c_2-1) \, [ (\tilde{q}_{,2} \, (\tilde{\omega}_{,2} - \tilde{\omega}) \, \tilde{r}_{,0,0})_{,1}  
      - \tilde{q}_{,2} \, (\tilde{\omega}_{,2} - \tilde{\omega}) \, \tilde{r}_{,0,0} ]   \, , \nonumber \\
     \tilde{q}_{,1} &=& (c_1-1)^{-1} \, [ (c_1 + \varphi_{,1} - \varphi) \, \tilde{q}_{,0} - q ]  \, , \nonumber \\
     \tilde{r}_{,-1} &=& (c_1-1)^{-1} \, [ \tilde{r}_{,-0} \, (c_1 + \varphi - \varphi_{,-1})_{,-0} - r ]  \, ,   
\eez
respectively,
\be
  \lefteqn{  c_2 \, \Big( ( g_{,2} \, g_{,0}^{-1} )_{,1} - g_{,2} \, g_{,0}^{-1} \Big) 
         - c_1 \, \Big( ( g_{,1} \, g_{,0}^{-1} )_{,2} - g_{,1} \, g_{,0}^{-1} \Big)  } \hspace{1cm} \nonumber \\
  &=& (c_2-1) \, [ ( \tilde{q}_{,2} \, (\tilde{\omega}_{,2} - \tilde{\omega}) \, \tilde{r}_{,0} )_{,1}
     - \tilde{q}_{,2} \, (\tilde{\omega}_{,2} - \tilde{\omega}) \, \tilde{r}_{,0} ]  \, , \nonumber \\
  \tilde{q}_{,1} &=& (c_1-1)^{-1} \, [ c_1 \, g_{,1} \, g_{,0}^{-1} \, \tilde{q}_{,0} - \tilde{q} ] \, , \nonumber \\
  \tilde{r}_{,-1} &=& (c_1-1)^{-1} \, [ c_1 \, \tilde{r}_{,-0} \, g_{,-0} \, g_{,-1}^{-1} - \tilde{r} ] \, .
  \label{dKP_scs_g}
\ee

\paragraph{Scalar discrete KP equation with self-consistent sources.}
We consider the scalar case $m=1$ with $\tilde{\omega}_{,1} = \tilde{\omega}$. Writing 
\bez
    g = \frac{\tau_{,-0}}{\tau} \, , \qquad
    \tilde{q} = \frac{\rho_{,-0}}{\tau} \, , \qquad
    \tilde{r} = \frac{\sigma}{\tau_{,-0}} \, , 
\eez
the equations for $\tilde{q}$ and $\tilde{r}$ in (\ref{dKP_scs_g}) read
\be
    (c_1-1)\, \tau_{,0} \, \rho_{,1} + \tau_{,0,1} \, \rho - c_1 \, \tau_{,1} \, \rho_{,0} = 0 \, , \qquad
    (c_1-1)\, \tau_{,1} \, \sigma_{,0} + \tau \, \sigma_{,0,1} - c_1 \, \tau_{,0} \, \sigma_{,1} = 0 \, .
      \label{discrKPscs_1}
\ee
By using these equations, and choosing $\tilde{\omega}_{,2} - \tilde{\omega} =: K \, (c_1-1)/[c_1 (c_2-1)]$ 
to be constant, 
the first of equations (\ref{dKP_scs_g}) can be cast into the form
\bez
  \lefteqn{ \frac{1}{ \tau_{,0} \, \tau_{,1,2} } \Big( {c_2 \, \tau_{,0,1} \, \tau_{,2} - c_1 \, \tau_{,0,2} \, \tau_{,1}}
  - K \, \rho_{,2} \, \sigma_{,0,1} \Big) } \hspace{2cm} \\
  &=& \Big[ \frac{1}{ \tau_{,0} \, \tau_{,1,2} } \Big( {c_2 \, \tau_{,0,1} \, \tau_{,2} - c_1 \, \tau_{,0,2} \, \tau_{,1}}
   - K \, \rho_{,2} \, \sigma_{,0,1} \Big) \Big]_{,-0}  \, ,
\eez
which implies
\be
    c_2 \, \tau_{,0,1} \, \tau_{,2}  - c_1 \, \tau_{,0,2} \, \tau_{,1} - c_{12} \, \tau_{,0} \, \tau_{,1,2}
    = K \, \rho_{,2} \, \sigma_{,0,1} \, ,   \label{discrKPscs_2}
\ee
with an arbitrary scalar $c_{12}$ that does not depend on the discrete variable $k_0$.
Up to differences in notation, the system (\ref{discrKPscs_1}) and (\ref{discrKPscs_2}) coincides with 
the discrete KP equation (Hirota bilinear difference or Hirota-Miwa equation) 
with self-consistent sources considered in \cite{Doli+Lin14} (also see \cite{Hu+Wang06}). 
(\ref{dKP_scs_g}) thus constitutes a ``non-commutative'' generalization of the latter.

\paragraph{Some explicit solutions for vanishing seed.}
We set $\varphi_0 =0$ and $g_0=I$. Then the linear equations for $\theta$ and $\eta$ are satisfied by
\bez
  && \theta = (1-c_1^{-1})^{-k_1} \, (1-c_2^{-1})^{-k_2} \, a \, P^{k_0} \, P_1^{k_1} \,  P_2^{k_2} \, A \, , \qquad 
     P_i := P - c_i^{-1} \, I \, ,  \quad i=1,2,  \\
  && \eta = (1-c_1^{-1})^{k_1} \, (1-c_2^{-1})^{k_2} \, B \, Q^{-k_0} \, Q_1^{-k_1} \, Q_2^{-k_2} \, b 
              \, ,\qquad 
              Q_i :=  Q - c_i^{-1} \, I \, ,  \quad i=1,2,
    \eez
with constant matrices $a$, $b$, $A$, $B$, $P$ and $Q$. A corresponding solution of the equations 
for ${\tilde\Omega}$ is
\bez
    \tilde{\Omega} = \tilde{\omega} + B \, Q^{-k_0} \, Q_1^{-k_1} \, Q_2^{-k_2} \, 
       X \, P^{k_0} \, P_1^{k_1} \, P_2^{k_2} \, A \, ,
\eez
with a constant matrix $X$ that satisfies the Sylvester equation
\bez
    Q^{-1} \, X \, P - X = b \, a \, .
\eez
(\ref{dKP_solutions}) together with $g = I - \theta \, (\tilde{\Omega}^+)^{-1} \, \eta$ now
provides us with explicit solutions of the above matrix discrete KP equations with self-consistent sources.

\section{Conclusions}
\label{sec:conclusions}
\setcounter{equation}{0}
The present work clarifies the origin of self-consistent source extensions of integrable equations 
from the binary Darboux transformation perspective. The essential point is a deformation of the 
potential $\Omega$ that is central in this solution generating method. We presented an abstraction  
of the underlying structure in the framework of bidifferential calculus. Choosing realizations of 
the bidifferential calculus then leads to self-consistent source extensions of various integrable equations, 
and in this work we provided a number of examples, recovering known examples and obtaining generalizations  
to matrix versions. All this is not at last a demonstration of the power of bidifferential calculus. 
Generalizing an integrability feature of an integrable system to this framework opens the door toward 
a large set of integrable systems sharing this feature. It therefore establishes such a feature as a common 
property of a wide class of integrable systems and provides a universal proof. 

Our approach also demonstrated that self-consistent source extensions of integrable equations 
typically\footnote{Exceptions are given by (\ref{KdV_scs2}) and (\ref{Bouss_scs_2}), since here $\omega$ is constant. }
admit classes of solutions that depend on arbitrary functions of a single independent variable (which, 
of course, may be a combination of the independent variables used). 
We note that the ``source-generation method'' in \cite{Hu+Wang06} essentially consists of promoting 
constant parameters in soliton solutions of an integrable equation to arbitrary functions of a single variable. 

In Appendix~\ref{app:pKPhier}, we also showed that the (2+1)-dimensional generalization of the Yajima-Oikawa 
equation (\ref{2+1YO}) belongs to the class of systems addressed in this work. As a consequence, its soliton solutions 
involve arbitrary functions of a combination of independent variables.

\vskip.2cm
\noindent
\textbf{Acknowledgments.}  
The authors are grateful for discussions with Adam Doliwa and Maxim Pavlov. 
F M-H also thanks Xing-Biao Hu for a useful discussion. 
O.C. has been supported by an Alexander von Humboldt fellowship for postdoctoral researchers.

\renewcommand{\theequation} {\Alph{section}.\arabic{equation}}
\renewcommand{\thesection} {\Alph{section}}

\makeatletter
\newcommand\appendix@section[1]{
  \refstepcounter{section}
  \orig@section*{Appendix \@Alph\c@section: #1}
  \addcontentsline{toc}{section}{Appendix \@Alph\c@section: #1}
}
\let\orig@section\section
\g@addto@macro\appendix{\let\section\appendix@section}
\makeatother

\begin{appendix}

\section{The extended matrix pKP hierarchy} 
\label{app:pKPhier}
On the algebra $\cA_0$ of smooth functions of variables $x$ and $\bsy{t} = (t_1,t_2,\ldots)$ we define 
\be
    \d f = [\cE_{\mu_1} , f ] \, \xi_1 + [\cE_{\mu_2} , f] \, \xi_2 \, , \quad
   \bd f = [(\mu_1^{-1} - \pa) \cE_{\mu_1} , f] \, \xi_1 + [(\mu_2^{-1} -\pa) \cE_{\mu_2} , f ] \, \xi_2 \, ,
           \label{pKP_bidiff}
\ee
where $\pa$ is the partial derivative operator with respect to $x$, $\mu_1$ and $\mu_2$ are indeterminates, 
and $\cE_\mu$ is the Miwa shift operator, hence $\cE_\mu f = f_{[\mu]} \, \cE_\mu$, where 
$[\mu] := (\mu,\mu^2/2,\mu^3/3,\ldots)$ and $f_{[\mu]}(x,\bsy{t}) = f(x,\bsy{t}+[\mu])$ (see, e.g., \cite{DMH06func} 
and references cited there). We will 
also use the notation $f_{-[\mu]}(x,\bsy{t}) := f(x,\bsy{t}-[\mu])$. Let $\cA$ be the algebra $\cA_0$ extended 
by $\pa$, the Miwa shift operators, and their inverses. The maps $\d$ and $\bd$ extend to $\cA$ and matrices over 
$\bsy{\Omega} = \cA \otimes \bsy{\Lambda}$. 
In the following we choose
\bez
    \Delta = \Gamma = - I_n \pa \, , \qquad \kappa = \lambda = 0 \, .
\eez
(\ref{Delta,lambda,Gamma,kappa_eqs}) is then satisfied. The second equation in (\ref{c,gamma_in_terms_of_omega}) leads to
\bez
      \gamma = \left( \mu_1^{-1} (\omega_{[\mu_1]} - \omega) - \omega_{[\mu_1],x} \right) \, \cE_{\mu_1} \, \xi_1 
               + \left( \mu_2^{-1} (\omega_{[\mu_2]} - \omega) - \omega_{[\mu_2],x} \right) \, \cE_{\mu_2} \, \xi_2 \, .
\eez
We require again (\ref{G.o=o.D}), which means
\bez
        \omega_x =0 \, .
\eez
The system collected in Remark~\ref{rem:phi,q,r,hatOmega_sys} now consists of  
\be
  &&  (\mu_2^{-1}-\phi+\phi_{-[\mu_2]})_{-[\mu_1]} \, ( \mu_1^{-1} - \phi + \phi_{-[\mu_1]} )
  - (\mu_1^{-1}-\phi+\phi_{-[\mu_1]})_{-[\mu_2]} \, ( \mu_2^{-1} - \phi + \phi_{-[\mu_2]} )  \nonumber \\
  &&  - (\phi_{-[\mu_1]} - \phi_{-[\mu_2]})_x  \nonumber \\
  &=& \mu_1^{-1} q_{-[\mu_1]} \, (\omega - \omega_{-[\mu_1]}) \, r 
    - \mu_1^{-1} q_{-[\mu_1]-[\mu_2]} \, (\omega - \omega_{-[\mu_1]})_{-[\mu_2]} \, r_{-[\mu_2]} \nonumber \\
  &&  - \mu_2^{-1} q_{-[\mu_2]} \, (\omega - \omega_{-[\mu_2]}) \, r 
    + \mu_2^{-1} q_{-[\mu_2]-[\mu_1]} \, (\omega - \omega_{-[\mu_2]})_{-[\mu_1]} \, r_{-[\mu_1]} \, ,
              \label{pKPhier_scs_phi}
\ee
\be
    \mu_1^{-1} (q - q_{-[\mu_1]}) - q_x &=& (\phi - \phi_{-[\mu_1]}) \, q
        +  \mu_1^{-1} q_{-[\mu_1]} \, (\omega - \omega_{-[\mu_1]}) \, \hat{\Omega} \, , \nonumber  \\
    \mu_1^{-1} (r_{[\mu_1]} - r) - r_x &=& - r\, (\phi_{[\mu_1]} - \phi) 
        + \mu_1^{-1} \hat{\Omega} \, (\omega_{[\mu_1]} - \omega) \, r_{[\mu_1]} \, ,    \label{pKPhier_scs_q,r}
\ee
and 
\be
    \hat{\Omega}_x = - r \, q \, , \qquad
    \mu_1^{-1} ( \hat{\Omega}_{[\mu_1]} - \hat{\Omega} ) 
         = - r \, q_{[\mu_1]} + \mu_1^{-1} \hat{\Omega} \, (\omega_{[\mu_1]} - \omega) \, \hat{\Omega} \, .  \label{pKPhier_hatOmega_eqs}
\ee
Expansion of these equations in powers of the indeterminates $\mu_1$ and $\mu_2$ generates the deformed hierarchy equations. 
Taking $\mu_2 \to 0$ in (\ref{pKPhier_scs_phi}) leads to
\bez
    ( \phi_{t_1} - \phi_x - q \, \omega_{t_1} r )_{[\mu_1]} - ( \phi_{t_1} - \phi_x - q \, \omega_{t_1} r ) = 0 \, ,
\eez
which implies
\be
       ( \phi_{t_1} - \phi_x - q \, \omega_{t_1} r )_{t_i} = 0 \qquad \quad i=1,2,\ldots  \label{pKPhier_scs_cons}
\ee
Without the deformation, i.e., if $\omega=0$, we are forced to identify $t_1$ with $x$. (\ref{pKPhier_scs_phi}) then  
generates the pKP hierarchy, and (\ref{pKPhier_scs_q,r}) generates corresponding linear and adjoint linear systems.   
Now we observe that a non-vanishing 
$\omega_{t_1}$ requires $t_1$ and $x$ to be independent. Besides (\ref{pKPhier_scs_cons}), the first non-trivial equation obtained 
from (\ref{pKPhier_scs_phi}) appears at order $\mu_1^2 \mu_2$ (or, equivalently, $\mu_1 \mu_2^2$). It is an extension 
of the pKP equation. 
Expansion of (\ref{pKPhier_scs_q,r}) yields to lowest orders
\bez
  \mu_1^0: &&    q_{t_1} = q_x + q \, \omega_{t_1} \hat{\Omega} \, , \quad
                   r_{t_1} = r_x + \hat{\Omega} \, \omega_{t_1} r  \, ,  \\
  \mu_1^1: &&  q_{t_2} = q_{t_1t_1} + 2 \, \phi_{t_1} q 
                + ( q \, \omega_{t_1t_1} - q \, \omega_{t_2} - 2 \, q_{t_1} \omega_{t_1} ) \, \hat{\Omega}  \\
             && \quad \; \, = q_{xx} + 2 \, ( \phi_{t_1} - q \, \omega_{t_1} r ) \, q  + q \, \omega_{t_2} \, \hat{\Omega}  \, , \\
             && r_{t_2} = -r_{t_1t_1} - 2 r \, \phi_{t_1} 
                + \hat{\Omega} \, ( \omega_{t_1t_1} r + \omega_{t_2} r + 2 \, \omega_{t_1} r_{t_1} ) \\
             && \quad \; \, = -r_{xx} - 2 r \, ( \phi_{t_1} - q \, \omega_{t_1} r ) + \hat{\Omega} \, \omega_{t_2} r  \, , \\
  \mu_1^2: && q_{t_3} = q_{xxx} + \frac{3}{2} \, ( \phi_{t_2} + \phi_{t_1t_1} )\, q 
                + 3 \, \phi_{t_1} q_x - \frac{3}{2} \, q \, (\omega_{t_2} + \omega_{t_1t_1} ) \, r \, q 
                - 3 \, ( q \, \omega_{t_1} r \, q)_x \\
             && \qquad \;  - 3 \, q \, \omega_{t_1} \hat{\Omega} \, \omega_{t_1} r \, q  + q \, \omega_{t_3} \, \hat{\Omega}  \, ,  \\
             && r_{t_3} = r_{xxx} - \frac{3}{2} \, r \, ( \phi_{t_2} - \phi_{t_1t_1} ) + 3 \, r_x \phi_{t_1}
                + \frac{3}{2} \, r \, q \, ( \omega_{t_2} - \omega_{t_1t_1} ) \, r  \\
             && \qquad \;   - 3 \, ( r \, q \, \omega_{t_1} r )_x  - 3 \, r \, q \, \omega_{t_1} \hat{\Omega} \, \omega_{t_1} r  
                + \hat{\Omega} \, \omega_{t_3} r  \, , 
\eez
where we used from (\ref{pKPhier_hatOmega_eqs}) the first equation and $\hat{\Omega}_{t_1} = - r \, q + \hat{\Omega} \, \omega_{t_1} \, \hat{\Omega}$.

\begin{remark}
Via expansion of (\ref{pKP_bidiff}) in the indeterminates $\mu_1$ and $\mu_2$, one obtains in particular the 
bidifferential calculus given by 
\bez
     \d f = [\pa_{t_1} , f] \, \xi_2 \, , \qquad
     \bd f = [\pa_{t_1} - \pa , f ] \, \xi_1 + [\frac{1}{2} (\pa_{t_2} + \pa_{t_1}^2) - \pa_{t_1} \pa , f ] \, \xi_2 \, ,
\eez
which underlies the above evolution equations with variables $t_1$ and $t_2$. They will be considered in the next example.
\end{remark}

\begin{example} 
\label{ex:2+1YO}
Setting $\omega_{t_2} = 0$, the equations for $q_{t_2}$ and $r_{t_2}$ do not involve $\hat{\Omega}$.
Disregarding the equations for $q_{t_1}$ and $r_{t_1}$, in terms of\footnote{As a consequence of (\ref{pKPhier_scs_cons}), 
we have $u = -2 \phi_x + a$, where $a$ is an arbitrary $m \times m$ matrix only allowed to depend on $x$.} 
$u := - 2 \, (\phi_{t_1} - q \, \omega_{t_1} r)$, we obtain the system\footnote{Introducing the new independent 
variable $x' = x+t_1$, this simplifies to $u_{t_1} = - 2 \, (q \, \omega_{t_1} r)_{x'}$, $q_{t_2} 
 = q_{x'x'} - u \, q$, $r_{t_2} = -r_{x'x'} + r \, u$. }  
\be
    u_{t_1} - u_x = - 2 \, (q \, \omega_{t_1} r)_x \, , \qquad
    q_{t_2} = q_{xx} - u \, q  \, , \qquad
    r_{t_2} = -r_{xx} + r \, u \, ,   \label{preYOsys}
\ee
where the first equation results from (\ref{pKPhier_scs_cons}) for $i=1$. 
Via $t_2 \mapsto -\imag \, t_2$, and with the reduction $r=q^\dagger$ and $u,\omega_{t_1}$ Hermitian, 
the system becomes \cite{Grim77} 
\bez
    u_{t_1} - u_x = - 2 \, (q \, \omega_{t_1} q^\dagger)_x \, , \qquad  
    \imag \, q_{t_2} = q_{xx} - u \, q   \, .
\eez
The change of variables $x = x'-t_1'+t_2'$, $t_1 = t_1'-t_2'$, $t_2 = t_2'$, turns this into
\be
    u_{t_1'} = - 2 \, (q \, \omega_{t_1'} q^\dagger)_{x'} \, ,  \qquad 
    \imag \, ( q_{t_1'} + q_{t_2'} ) = q_{x'x'} - u \, q  \, ,         \label{2+1YO}
\ee 
which is a (2+1)-dimensional generalization \cite{Mel'83,Shul84,OOF89,Macc96,SSS99,Berk+Sido02,OMO07,CCFM15} 
of the \emph{Yajima-Oikawa} system 
\cite{Yaji+Oika76}. We note that $\omega_{t_1'}$ can be absorbed by a redefinition of $q$. 
Our derivation of this system via the above deformation of the pKP hierarchy is new, according to our 
knowledge.\footnote{It is well-known that the (1+1)-dimensional Yajima-Oikawa system arises from the KP hierarchy 
via a symmetry constraint \cite{KSS91,Stra96,Oeve+Cari98}. It still has to be clarified how this is related to the 
above deformation of the pKP hierarchy. }
\end{example}

Adressing our solution generating method, the linear equations (\ref{theta_eq}) and (\ref{eta_eq}) read
\be
    \mu_1^{-1} \, (\theta - \theta_{-[\mu_1]}) - \theta_x = (\phi_0 - \phi_{0,-[\mu_1]}) \, \theta \, , \qquad
    \mu_1^{-1} \, (\eta_{[\mu_1]} - \eta) - \eta_x = - \eta \, (\phi_{0,[\mu_1]} - \phi_0) \, ,
       \label{pKPhier_scs_lin_eqs}
\ee
and (\ref{scs_Omega_eqs}) yields 
\be
    \Omega_x = - \eta \, \theta  \, , \qquad
    \mu_1^{-1} (\Omega_{[\mu_1]} - \Omega) = - \eta \, \theta_{[\mu_1]} + \mu_1^{-1} (\omega_{[\mu_1]}-\omega) \, .
       \label{pKPhier_Omega}
\ee
Recall that $\hat{\Omega} = - \Omega^{-1}$. Solutions of the system (\ref{pKPhier_scs_phi}), (\ref{pKPhier_scs_q,r}) and (\ref{pKPhier_hatOmega_eqs})
are then obtained via (\ref{phi,q,r}).

\begin{example} 
Solutions of the linear equations 
$\theta_{t_1} = \theta_x$, $\theta_{t_2} = \theta_{xx}$, $\eta_{t_1} = \eta_x$, $\eta_{t_2} = -\eta_{xx}$,  
obtained from (\ref{pKPhier_scs_lin_eqs}) with constant seed $\phi_0$, are given by 
\bez
    \theta = a \, e^{\vartheta(P)} \, A \, , \quad 
    \eta = B \, e^{-\vartheta(Q)} \, b \, , \qquad \vartheta(P) = P \, (x+t_1) + P^2 t_2 \, ,
\eez
with constant matrices $a,b,A,B,P,Q$ of appropriate size. The corresponding solution of (\ref{scs_Omega_eqs}) is
\bez
     \Omega = B \, e^{-\vartheta(Q)} \, X \, e^{\vartheta(P)} \, A + \omega(t_1) \, , \qquad
     Q \, X - X \, P = b \, a \, .
\eez
Solutions of the above system (\ref{preYOsys}) are then given by 
\bez
    u = 2 \, (\theta \, \Omega^{-1} \eta)_x \, , \quad
    q = \theta \, \Omega^{-1} \, , \quad 
    r = \Omega^{-1} \eta \, .
\eez
We would rather like to obtain solutions of (\ref{preYOsys}) with $\omega_{t_1}$ replaced by $-I_n$, since in this case   
(\ref{2+1YO}) becomes the system studied, e.g., in \cite{OOF89}. Since this can be achieved by a transformation of $q$ and $r$,  
this means that, for \emph{any} matrix function $\omega(t_1)$, we obtain a class of solutions. 

In the simplest case, where $m=n=1$, let us set $a=b=A=B=1$ and $Q=-P^\dagger$. We write $p=\mathrm{Re}(P)$, $s = \mathrm{Im}(P)$, 
and $\omega = - e^{2 \, \alpha}/( 2 \, p )$, with a real function $\alpha(t_1)$. Then we obtain 
\bez
   q = - p \, e^{- \alpha + \imag \, \delta} \mathrm{sech}(\beta)  \, , \qquad 
   u = - 2 \, p^2 \, \mathrm{sech}^2(\beta)  \, ,
\eez
with $\beta = p \, (x+t_1) - \alpha(t_1) + 2 \, s \, p \, t_2$ and $\delta = s \, (x+t_1) - (p^2 - s^2) \, t_2 + \delta_0$, 
where $\delta_0$ is an arbitrary real constant. 
Then $\tilde{q} = \sqrt{-\omega_{t_1}} \, q = - \sqrt{ p \, \alpha_{t_1}} \, e^{\imag \, \delta} \, \mathrm{sech}(\beta)$, 
where $p \, \alpha_{t_1} >0$, and the above $u$ solve the system
\bez
  u_{t_1} - u_x - 2 \, (|\tilde{q}|^2)_x =0 \, , \qquad 
  \imag \, \tilde{q}_{t_2} = \tilde{q}_{xx} - u \, \tilde{q}  \, ,
\eez  
which translates into the (2+1)-dimensional Yajima-Oikawa system (\ref{2+1YO}) via the transformation of variables stated in 
Example~\ref{ex:2+1YO}. The above solution then becomes the 1-soliton solution, obtained previously in \cite{OOF89}, 
if we choose $\alpha = k_1 \, t_1 + k_0$, with real constants $k_0,k_1$. We learned, however, that $\alpha$ is allowed to 
be an \emph{arbitrary} function of $t_1$. More generally, we obtain multi-soliton solutions depending on several 
arbitrary functions of $t_1$. 
\end{example}

\paragraph{Reduction $t_1 =x$.} This requires $\omega_{t_1}=0$. The first of equations (\ref{pKPhier_Omega}) is then redundant. 
With this reduction, we recover (\ref{pKP_scs}), (\ref{pKP_q-sys_scs}) and (\ref{pKP_r-sys_scs}) by expansion of 
(\ref{pKPhier_scs_phi}) and (\ref{pKPhier_scs_q,r}). The next equations in the deformed hierarchy are
\bez
  &&  3 \, \phi_{x t_4} - 2 \, \phi_{y t} - \phi_{xxxy} - 3 \, (\phi_x{}^2)_y - 3 \, \{\phi_y, \phi_{xx}\}
    + [ \phi_x , 4 \, \phi_t - \phi_{xxx}] \\
  &=& 3 \, (q \, \omega_{t_4} r)_x + 2 \, (q \, \omega_t r_x - q_x \omega_t r)_x - 2 \, (q \, \omega_y r)_t
    + \frac{3}{2} (q \, \omega_y r_y - q_y \omega_y r)_x \\ 
  &&  - 3 \, (q_x \omega_y r_x)_x + \frac{1}{2} (q \, \omega_y r)_{xxx} \, ,
\eez
where $y=t_2$ and $t=t_3$, and
\bez
    q_{t_4} &=& q_{xxxx} + 4 \, \phi_x q_{xx} + 2 \, (\phi_y + 2 \, \phi_{xx}) \, q_x 
                + ( \frac{4}{3} \phi_t + \phi_{xy} + \frac{5}{3} \phi_{xxx} + 2 \, \phi_x{}^2 ) \, q \\
            &&  - 2 \, (q \, \omega_y r) \, q_x - (q \, \omega_y r)_x \, q - \frac{4}{3} \, (q \, \omega_t r) \, q
                - q \, \omega_{t_4} \Omega^{-1}   \, , \\
    r_{t_4} &=& - r_{xxxx} - 4 \, r_{xx} \phi_x + 2 \, r_x (\phi_y-2 \, \phi_{xx})
                - r \, ( \frac{4}{3} \phi_t - \phi_{xy} + \frac{5}{3} \phi_{xxx} + 2 \, \phi_x{}^2 ) \\
            &&  - 2 \, r_x (q \, \omega_y r) - r \, (q \, \omega_y r)_x +  \frac{4}{3} r \, (q \, \omega_t r)
                - \Omega^{-1} \omega_{t_4} r \, .
\eez
The corresponding additional equation for $\Omega$, obtained from (\ref{pKPhier_Omega}), is
\bez
    \Omega_{t_4} = - \eta \, \theta_{xxx} + \eta_x \theta_{xx} - \eta_{xx} \theta_x + \eta_{xxx} \theta
                   - 4 \, \eta \, \phi_{0,x} \theta_x + 4 \, \eta_x \phi_{0,x} \theta 
                   - 2 \, \eta \, \phi_{0,y} \theta + \omega_{t_4} \, .
\eez
\end{appendix}

\end{document}